\documentclass[sigconf]{acmart}
\pdfoutput=1

\usepackage[ruled,vlined]{algorithm2e} 
\usepackage[noend]{algpseudocode}
\usepackage{multirow}
\usepackage{subfigure}
\usepackage{nicefrac}

\usepackage{xspace}
\usepackage{color}
\usepackage{graphicx}
\usepackage{url}
\usepackage{xspace}

\newcounter{todocounter}

\makeatletter
\algocf@newcommand{KwDataXX}[1]{%
  \sbox\algocf@inputbox{\hbox{\KwSty{Data}\algocf@typo: }}%
  \ifthenelse{\boolean{algocf@inoutnumbered}}{\relax}{\everypar={\relax}}%
  {\let\\\algocf@newinput\hspace{\wd\algocf@inputbox}\hangindent=\wd\algocf@inputbox\hangafter=\wd\algocf@inputbox#1\par}%
  \algocf@linesnumbered
}
 \let\oldnl\nl
\newcommand{\nonl}{\renewcommand{\nl}{\let\nl\oldnl}}
\makeatother

\usepackage{hyperref}

\hypersetup{
    colorlinks=true,
    linkcolor=red!40!black,
    citecolor=green!60!black,,
    filecolor=magenta,
    urlcolor=blue
}

\newtheorem{hypothesis}{Hypothesis}

\renewcommand\footnotetextcopyrightpermission[1]{} 
\begin{document}

\title{Learning Interesting Categorical Attributes\\ for Refined Data Exploration}

\titlenote{This work has been supported by the German Research Foundation (DFG) under grant MI 1794/1-1. This is the extended 
version of the short paper presented at EDBT'16~\cite{Pal16}.}

\author{Koninika Pal}
\affiliation{%
  \institution{TU Kaiserslautern}
  \city{Kaiserslautern}
  \country{Germany}
}
\email{pal@cs.uni-kl.de}

\author{Sebastian Michel}
\affiliation{%
  \institution{TU Kaiserslautern}
  \city{Kaiserslautern}
  \country{Germany}
}
\email{michel@cs.uni-kl.de}

\renewcommand{\shortauthors}{}

\begin{abstract}
This work proposes and evaluates a novel approach to determine interesting 
categorical attributes for lists of entities. 
Once identified, such categories are of immense value to allow constraining (filtering)
a current view of a user to subsets of entities. We show how a  classifier is trained
 that is able to tell whether or not a 
categorical attribute can act as a constraint, in the sense of human-perceived interestingness.
The training data is harnessed from Web tables, treating the presence or absence of a table as an indication 
that the attribute used as a filter constraint is reasonable or not. For learning the classification model, 
we review four well-known statistical measures (features) for categorical attributes---entropy, 
unalikeability, peculiarity, and coverage. We additionally propose 
three new statistical measures to capture the distribution of data, tailored to our main objective.
The learned model is evaluated by relevance assessments obtained through a user study, reflecting 
the applicability of the approach as a whole and, further, demonstrates the superiority of the proposed 
diversity measures over existing statistical measures like information entropy.
\end{abstract}

\maketitle

 
\section{Introduction}
\label{sec:intro}

 Rendering large and heterogeneous data accessible to users requires mechanisms that allow querying or exploring it without prior domain knowledge.
Consider for instance knowledge bases like YAGO, Freebase,  or DBPedia that alone contain already hundreds of millions of facts for tens of millions of entities 
from all kinds of domains and types.  One classical approach to make such vast amounts of information 
 accessible  to users is to 
organize data  into specific categories according to attributes, e.g., scientists born in Norway, capital cities located in Europe, the tallest buildings in USA, 
in order to constrain the view of data explorers to such subsets.  But, are  all possible categories insightful and, if not,  who defines meaningful categories?

Consider the case of large businesses with several retail stores across the USA. 
Clearly, data analysts are likely to investigate properties like the best selling items overall,
but in particular also the best selling items per state or city.
Likewise, it might also be interesting to investigate the sales of retail stores for specific product 
categories, or deals accomplished per employee. Analysts frequently use the drill-down operation
in OLAP~\cite{DBLP:journals/datamine/GrayCBLRVPP97,DBLP:conf/edbt/SarawagiAM98}  over predefined categories to analyze such cases. 
For a reasonably small scenarios with well-defined  schemata, telling which categories 
(dimensions) are interesting can be accomplished by domain experts, 
manually~\cite{DBLP:conf/icde/JoglekarGP16}. However, when turning our attention to 
arbitrary, per se unknown scenarios in the age of Big Data, heterogeneity, 
dynamics, and scale strongly advocate solely automated means. 

The overall task we tackle in this work is the following:
Given a (Web) table that contains entities and their attributes, we want to determine 
those categorical attributes (i.e., columns of the table)  that can act as filters to constrain the 
focus of the table, i.e., to categorize the entities, thus, providing more focused and comprehensible information to users.

Let us introduce the problem through an example. Table~\ref{fig:rank} is showing part of a 
Wikipedia table reporting on the world's tallest buildings, sorted by height. This list is quite long,
as {\it very many} tall buildings from many countries all over the world are captured.
A refined view of this large table can be defined by imposing a constraint on the attribute 
{\it country}, such as country=`United States' or country=`China'.  Browsing through such constrained tables
fosters  exploration/understanding  of datasets at hand and can further answer specific information needs of users.

\begin{table}[!t]
\caption{The World's Tallest Buildings (Wiki\-pedia)}
    \label{fig:rank}
\small
\begin{tabular}{|p{3.2cm}|l|l|l|} \hline
\multicolumn{1}{|c|}{\bf Building} & \multicolumn{1}{|c|}{\bf City} & \multicolumn{1}{|c|}{\bf Country}  &\multicolumn{1}{|c|}{\bf Height} \\ \hline \hline
Burj Khalifa & Dubai & UAE & 828m \\ \hline
Shanghai Tower & Shanghai &  China & 632m  \\ \hline
Abraj Al-Bait Clock Tower & Mecca & Saudi Arabia & 601m  \\ \hline
Ping An Finance Centre & Shenzhen & China & 599m  \\ \hline
Goldin Finance 117 & Tianjin & China & 596m  \\ \hline
One World Trade Center & NY City & United States & 541m  \\ \hline
\end{tabular}    
\end{table}

But are all attributes useful in the sense that they define interesting subsets? 
For humans with domain knowledge, it is a relatively simple task to decide whether a categorical attribute is interesting to be 
used for further categorizing the entity list, although sometimes 
subjective. Categorizing skyscrapers by continent, for instance, seems very reasonable, while for
organizing them by architect it depends on the number of skyscrapers per architect---boring, if each architect designed only one or two skyscrapers.
With large and heterogeneous data available, specifically on the 
Web, hiring domain experts annotating attributes manually is  infeasible. 
In this work, we propose a fully automated framework in order to learn a classification model 
that can identify  categorical attributes that are suitable for categorizing 
entities. Suitable in the sense of human-perceived interestingness. 

The human perception of `interestingness' is a complex concept that asserts {\it unexpectedness, 
conciseness, coverage, utility, and diversity}~\cite{Geng:2006:IMD:1132960.1132963}. 
 Hence, finding suitable statistical measures 
that capture  interesting or non-interesting characteristics in categorical attributes renders the learning problem
difficult. In this paper, we first investigate  existing objective measures of interestingness for 
categorical attributes~\cite{Geng:2006:IMD:1132960.1132963, journals/Gary07}. As we will see by anecdotal evidence and later also by experimental results, these 
empirical probability-based measures fail to capture all aspects of our task to identify 
interesting categories in many cases, as discussed in Section~\ref{sec:novel}. To address the identified shortcomings,
 we propose three novel statistical measures, \textsf{P-Diversity, P-Peculiarity, 
and Max-Info-Gap} for categorical attributes. Finally, we learn a robust and accurate 
classification model using support vector machine {\em(SVM)} over combinations of proposed 
and existing measures. By means of a  user study, we show
that the trained classifier is able to predict those categorical attribute that are suitable for further categorization 
of entities---in a human-perceived sense. We will also see later from experimental studies that 
the proposed statistical measures are more effective in capturing `interestingness' compared to commonly used measures like entropy or coverage.
To the best of our knowledge, this is the first full-fledged approach that enables the identification of meaningful categorical  attributes,
a generic and  widely applicable ingredient to data exploration and analytics.

\subsection{ Problem Statement and Notation}
\label{sec:prob_stat}

Our objective is to understand which categorical attribute of a specific entity-centric table
will be perceived suitable by humans for the task of defining a meaningful subset of the entities.
Hence, in this work, {\em a categorical attribute is considered interesting if it is suitable for further categorization of entities. }

In order to do so, we investigate a set of tables $\mathcal{R}$, where a table $r\in \mathcal{R}$ represents a list of 
entities of a specific type together with their attributes $\mathcal{A}$. 
A set of statistical measures $\mathcal{F}$ is used to map the categorical attributes 
(i.e., columns of tables) to the feature space, in order to train a classifier $\mathcal{C}$ that can predict which categorical attributes are 
interesting for categorizing the entities of the table. 

Consider again Table~\ref{fig:rank} and let us  denote with 
$\mathcal{V}_{\text{country}} = \{ \text{UAE (1)},$ $\text{China (3)}, \text{Saudi Arabia (1)}, \text{United States (1)}\}$ the set of values
for the attribute  $\text{country}$. The numbers in parentheses express the multiplicity.
Statistical measures can now be computed based on these numbers, for instance, the Shannon entropy of the according frequency distribution would be $1.792$.
If we knew that the attribute $\text{country}$ is an interesting attribute for categorizing tall buildings, then 
we could, roughly speaking, learn that an entropy around $1.792$ might be an indicator for interesting attributes, in any table we encounter.

To bring this toy example to larger scale, in order to build a reliable classifier, we face the following  {\bf main tasks}: 
\begin{itemize}
\item First, to the best of our knowledge, there is no dataset 
available in literature that provides information on which 
categorical attributes are useful for the categorization of the entities of a specific class.
Such training data is, however, crucial for training a classifier, thus, needs to be acquired first---and we want to do so without any manual human intervention.
\item Second, we have to identify suitable measures (statistical features) that can capture the characteristics of categorical 
attributes, tailored to our context of interestingness. Existing features have certain drawbacks that hinder them  grasping important characteristics.
We will review such weaknesses and explain how the newly proposed measures can reflect them. \\
\end{itemize} 
{\bf Notation:} 
An attribute $a \in \mathcal{A}$ of a table can be of three kinds: {\it (i)} the subject of the table, 
denoted as $a_s$, which represents the class of the entities, {\it (ii)} the measuring attributes, 
denoted as $a_{n} \in \mathcal{N}$, which is of numeric type and {\it (iii)} the categorical attributes, referred as $a_c$. 
Additionally, for each $r\in \mathcal{R}$, we also extract some metadata $M$ that holds the constraint and ranking criterion of 
the table, denoted as $r_{cons}$ and $r_{cr}$, respectively, if available. Following this notation, 
the table $r$ depicted in Table~\ref{fig:rank} is represented as 
$r= ( a_s, a_{n}, a_{c1}, a_{c2}, M )$, where $a_s$ = Building, $a_{n}$ = height,
$a_{c1}$ = city, and $a_{c2}$ = country. No constraints are extracted for this table. Each attribute $a\in \mathcal{A}$ is associated with a value set, denoted $\mathcal{V}_a$. 
An example of   $\mathcal{V}_a$ has been shown above.  

\subsection{Sketch of our Approach}
\label{sec:framework}

\begin{figure}[t]
    \centering    
   \includegraphics[width=.9\columnwidth]{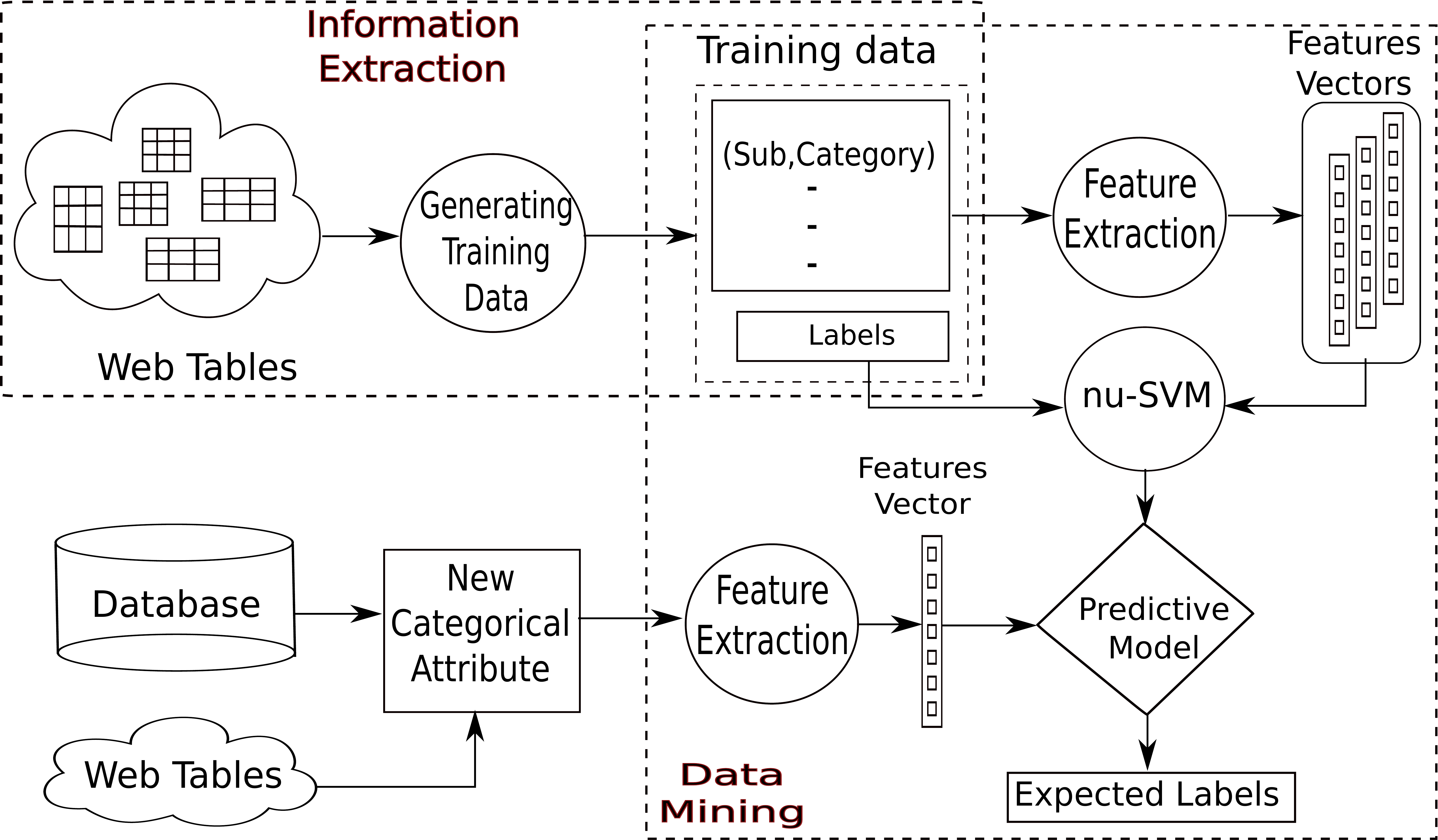}
      \caption{Framework of mining categorical attributes}
    \label{fig:framework}
\end{figure}

Our whole approach is divided into two main components as shown in Figure~\ref{fig:framework}. The {\em Information Extraction} component first creates the training samples
from Wikipedia tables in a completely automated manner. For this purpose, we identify the 
categorical attributes from a set of Wikipedia tables. Then, we find out
whether retrieved categorical attributes will be labeled as `interesting' or 
`non-interesting' for further categorization of the entities based on a 
central hypothesis, proposed in Section~\ref{sec:createdata}. 

After labeling the categorical attributes, the {\em Data Mining} component extracts the feature vector 
$\mathcal{F}$ for each categorical attribute in the training data. $\mathcal{F}$ comprises existing 
and newly proposed statistical measures. Then, a classifier $\mathcal{C}$ is trained over the extracted feature vectors 
using $\nu$-SVM. 

\subsection{Contribution and Organization}

With this work, we make the following contributions:
\begin{itemize}
\item We describe a framework to harness training samples of interesting and 
non-interesting categorical attributes from tables without explicit human interaction.
\item We investigate statistical measures that can capture the interestingness of 
categorical attributes and propose three new statistical measures tailored to our main objective.
\item We have conducted a comprehensive evaluation, including a user study, 
demonstrating the applicability of the general approach and the superiority of the newly proposed features.
\item The sample training data retrieved from Wikipedia tables, relevance assessments from user, and trained classifiers are made public.  
\end{itemize}

A preliminary version of this work has been published as a short paper at EDBT'16~\cite{Pal16}.

The paper is organized as follows. Section~\ref{sec:createdata} proposes the working hypothesis 
and the algorithm to extract training data. Section~\ref{sec:learning} discusses the proposed 
statistical measures and the existing ones, and introduces the learning 
model. Section~\ref{sec:experiments} presents the experimental results. An overview of related 
work is presented in Section~\ref{sec:relatedworks} and Section~\ref{sec:conclusion} concludes the paper.


\section{Automated Extraction of Training Data}
\label{sec:createdata}
 
In this section, we describe how  training data can be obtained in a fully automated fashion.
That means, for categorical attribute $a_c$ (i.e., a column of a table) that appears in a table 
for entity type $a_s$, we need to determine the label (interesting or not-interesting) that tells 
whether the attribute allows a suitable categorization of the entities, reflecting a human notion
of a meaningful categorization or not. But how can we determine the label without human effort? 

We put forward the working assumption that the presence and absence of Web tables is 
an indicator of general interest or disinterest of humans in such tables. 
And following this assumption,  the presence of a Web table makes 
a categorical attribute, that is used as a constraint to create that very table, interesting. 
This observation is cast into our general hypothesis given next.  We will see later by 
experiments on human relevance assessments that this hypothesis is in fact well-grounded, and discuss limitations below.

\begin{hypothesis}   A categorical attribute in a table is considered interesting,
thus its statistical features are positive training samples, iff we find at least one table over the same entity class that is created by imposing a constraint over that categorical attribute. 
 \label{def1} 
\end{hypothesis}

\begin{table}[t]
     \caption{List of Tallest Buildings in the Unites States}
    \label{fig:newrank}
\center
\small
\begin{tabular}{|p{3cm}|p{2.2cm}|p{1.85cm}|} \hline
\multicolumn{1}{|c|}{\bf Building} & \multicolumn{1}{|c|}{\bf City} &\multicolumn{1}{|c|}{\bf Height} \\ \hline \hline
One World Trade Center & New York City & 541m \\ \hline
Willis Tower & Chicago & 442m  \\ \hline
432 Park Avenue & New York &  426m \\ \hline
\end{tabular}
\end{table}

Let us walk through an {\bf example} to explain the intuition behind this hypothesis. In 
Table~\ref{fig:rank}, we observe that entities of class `building' are displayed, together with 
the categorical attributes `country' and  `city'. By browsing through Wikipedia, we also find 
another table, namely the
{\it List of Tallest Buildings in the United States}, shown in Table~\ref{fig:newrank}. As will be described in more detail below, we use in fact
the caption of the table, respectively the title of the corresponding Wikipedia table to determine the attribute used as constraint.
Clearly, both tables are created on the same entity class, building, and the constraint 
`country=United States' is applied in Table~\ref{fig:newrank}, we see this in the title/caption of that table. We also find that `United States' 
is one of the categorical value for attribute `country' in Table~\ref{fig:rank}.
Consequently, according to our hypothesis, `country' is considered interesting for entity 
class `building', as we found (at least) one table (Table~\ref{fig:newrank}) that is created by imposing 
the constraint `United States' on  the categorical attribute `country'  from Table~\ref{fig:rank}.
Note that the table ``List of Tallest buildings in United States" may not be a subset of the 
table shown in Table~\ref{fig:rank}, this is irrelevant for our task, however.

Now, once we have found such a pair of tables, we consider Table~\ref{fig:rank} as parent and Table~\ref{fig:newrank} as child. 
Then we extract statistical features (for instance,  information entropy) from the 
categorical attribute `country' of the parent table and consider it as a positive sample in
our training data. 

Although the final classifier is independent of specific entity types, while generating the training data from tables, an association between subject $a_s$ and 
categorical attribute $a_c$ is required, as a categorical attribute can be associated with many 
entity types (e.g., 'length' can be an attribute for highways, bridges, beaches, etc.), thus, the pair 
($a_s, a_c$) provides a unique identification for features retrieved for attribute $a_c$ for 
 entity class $a_s$. It would be misleading, or simply wrong, to search for {\it any} 
table (irrespective of matching entity type) that was generated by using a constraint on 
categorical attribute $a_c$, to conclude anything useful from the statistics computed from 
any table that has such an attribute $a_c$. The final classifier is independent of the 
entity class, as it operates solely on statistical measures retrieved from categorical attribute.

\subsection{Algorithm}
To find the parent-child relation between two tables, in order to retrieve the label for categorical 
attributes and its statistics, a brute-force method would visit all pairs of tables.
Avoiding the brute-force method, Algorithm~\ref{alg:1}  scans
the tables twice to retrieve the training samples: once to learn from table metadata which constrained tables exist for an entity type, and, once to draw the connections between table columns/attributes and existing constrained tables. More precisely, in the first phase, 
only the table header and metadata for all tables are scanned to
retrieve the constraint $r_{cons}$ and subject $a_s$ of the table. 
The detailed way of retrieving metadata and subject  is discussed 
in Section~\ref{sec:wiki}.
The extracted information is subsequently used to build an index in form of a simple map, called $cons\_map$. It takes the 
constraint $r_{cons}$ retrieved from a table as key and corresponding subject $a_s$ 
of that table as value. For example, considering the two tables {\it ``List of Tallest Buildings in 
the United States" and ``List of Universities in the United States"}, the constraint in both 
tables is {\it United State} and the corresponding entries in $cons\_map$ is as follows:
$$ \text{United State} \rightarrow \text{\{buildings, universities\}}. $$ 

\begin{algorithm}[t]
\DontPrintSemicolon
\LinesNumbered
\KwData{Initialization\;
$\mathit{cons\_map}: \{key:constraints, value: subjectList[])\}$\;
training samples: $\mathit{interesting}[]$ \tcp*{a list of $\{(a_s, a_c),\mathcal{F}\}$ where $a_c$ is used to categorize $a_s$.}
\nonl Negative training samples: $\mathit{nonInteresting}[]$ \tcp*{a list of $\{(a_s, a_c),\mathcal{F}\}$ where $a_c$ is not used to categorize $a_s$ \nonumber}}
\SetKwFunction{proc}{generateSamples}
      \SetKwProg{myproc}{Procedure}{}{}
  \myproc{\proc{Web tables $\mathcal{R}$}}{	 
 \tcc{Scan on $\mathcal{R}$ to build $\mathit{cons\_map}$}
	\For{$r \in \mathcal{R}$}{ 
		$r.a_s , r_{cons} \gets parse\_metadata(r)$ \; \label{alg:line6}
		add $(r.a_s)$ to the list of $\mathit{cons\_map}[r_{cons}]$}
\tcc{Scanning $\mathcal{R}$ to build training samples}	 
	\For{$r \in \mathcal{R}$}{
	 	$List\{r.A\} \gets parse(r)$ \tcp*{Parsing all colums}
	 	$List\{r.A_c\} \gets List\{r.A\} \setminus (r.a_s \cup \mathcal{N})$\label{alg:line11}  \tcp*{Removing numeric attributes}
		\For{$a_c \in List\{r.A_c\} $}{
		$\mathcal{F} \gets calculateFeatures(\mathcal{V}_{a_c})$\;
		\tcc{Find existence of parent, child table in $\mathcal{R}$ based on $\mathcal{V}_{a_c}$}
			\For{$x \in \mathcal{V}_{a_c}$}{
			 	$subjectList \gets \mathit{cons\_map}[x]$ \label{alg:line14} \;
				\eIf{$r.a_s \in subjectList $}{
					add  $\{(r.a_s, a_c), \mathcal{F}\}$ to $\mathit{interesting}[]$\;
					 \text{break}\;}
					{ add $ \{(r.a_s, a_c), \mathcal{F}\}$ to $\mathit{noninteresting[]}$\;}}}}
\KwRet{$\mathit{interesting[]}, \mathit{noninteresting[]}$}\;}	
\caption{Generating Training Samples}
\label{alg:1}
\end{algorithm}

After creation of $cons\_map$, a second complete scan over the tables and their columns and headers is done to
retrieve all the attributes and their positive or negative label. We remove numeric attributes
(cf., Line~\ref{alg:line11} in Algorithm~\ref{alg:1}) and only consider the categorical 
attributes for a specific entity class, i.e., pairs of $(a_s, a_c)$, from each Web table, in order to create the 
training samples and generate automatically the labels based on Hypothesis~\ref{def1}.

According to this main hypothesis, the 
label of a categorical attribute is found by identifying the parent-child relation between  
tables, using that categorical attribute. To determine this, we take
$V_{a_c}$, associated with a categorical attribute $a_c$ for a specific entity type $a_s$, and
check the $cons\_map$ for each categorical value in $V_{a_c}$ until we find a match. 
Once a value from $V_{a_c}$ is contained as key in  $cons\_map$, we scan the subject list 
associated with the key to see  whether the entity class $a_s$ is contained in the 
list. If it is, we know that there exists  at least one table that 
is built over same entity class $a_s$  and is using one of the categorical value from $V_{a_c}$ 
associated with attribute $a_c$ from the current table we are scanning. Clearly, we identify 
the current table as  a parent table and also find the existence of a child table 
using $a_c$. Hence, we label the pair $(a_s, a_c)$ as ``interesting" based on our hypothesis 
and  consider it a positive training samples. If we did not find any match with the key in $cons\_map$ for any of the value from  $V_{a_c}$, 
we know  that no child table is found based on the attribute $a_c$. Hence, the pair $(a_s, a_c)$ is labeled  as 
``non-interesting'' and is, thus,  considered a negative training sample. 

While retrieving $V_{a_c}$ for a categorical attribute $a_c$, the frequency count of the entities appearing in the table 
associated with each categorical value in $\mathcal{V}_{a_c}$ is also captured 
(e.g., in Table~\ref{fig:newrank}, $\mathcal{V}_{city}=\{\text{New York City } (2), \text{Chicago }(2)\}$). 
This information is then used to map a pair 
$(a_s, a_c)$ to feature space $\mathcal{F}$,  capturing the empirical characteristics, such as information entropy, 
of the pair $(a_s, a_c)$; discussed in detail in Section~\ref{sec:learning}. 

Note that Web tables are not always fully consistent in data representation and 
column descriptions~\cite{DBLP:conf/kdd/SarawagiC14,DBLP:conf/cikm/IbrahimRW16}. 
As ambiguous representations of numeric types can generate wrong classification labels to 
categorical attributes, Algorithm~\ref{alg:1} excludes all attributes of numeric type, which 
results in removing categorical attributes such as years etc., when generating the training data.
 
\subsection{Harnessing Wiki\-pedia}
\label{sec:wiki}

In this work, we specifically use the English Wiki\-pedia corpus to generate  training samples.
The major difficulties in extracting and understanding information from Web tables arise due 
to inherent heterogeneity in schema and data representation~\cite{DBLP:conf/wsdm/CrestanP11, DBLP:conf/er/WangWWZ12}. 
Here, we discuss in more detail why Wiki\-pedia is an excellent source for our endeavor.

First of all, by enforcing collaborative editing policies and controlling duplicated information across
multiple pages, Wiki\-pedia maintains high information quality and thus {\it generally 
considered credible for knowledge exploration}~\cite{DBLP:journals/pvldb/ChirigatiLKWYZ16, DBLP:conf/cikm/MilneW08, DBLP:conf/kdd/BhagavatulaND13}. 
We also excluded  user pages  to avoid biased data. 

Second, Wiki\-pedia contains tables that provide surprisingly many categorical attributes.
Such information is not available in Web portals like \href{http://rankopedia.com}
{rankopedia.com} or \href{http://ranker.com}{ranker.com} where tables are created based on crowdsourcing, with only numeric attributes (mainly number of upvotes) being available for the entities. 
More precisely,  $3/4$ of all Wiki\-pedia tables investigated for this work  contain categorical attributes.
 The distribution of the number of categorical  attributes per table can be well described by a Poisson distribution with mean $\lambda=1.9$ 
(with relative sum squared error of $<0.00001$). 

Third, the structure of the tables in Wiki\-pedia is quite consistent and the metadata of tables, required by Algorithm~\ref{alg:1}, can be extracted.
However, we noticed that often enough not all pages have sufficient information associated via html tags, not even the title of the table. 
Hence, retrieving  metadata from arbitrary tables would require sophisticated NLP techniques and perhaps  
further demand user interaction for checking the correctness of the results. Thus, in this work, we 
consider only those tables that are created with a page title beginning with the phrase 
``List of $\ldots$" and have the property of being sortable. This greatly helps to accurately collect 
metadata, such as subject, constraint, etc., of the tables by parsing the title/caption of the 
table or the title of the Wiki\-page. These  page titles have a very simple 
sentence structures that can be easily parsed by using {\it propositions from the English 
dictionary}, in order to retrieve the subject and the constraints of a table. 
For example, from the page  title ``List of Tallest Buildings in the Unites States", we retrieve subject of the table as `Tallest 
Buildings' and `Unites States' as constraint, based on prepositions `of' and `in'. Although 
sometimes page titles are more complex than the example given, they are still easily parseable 
as usually much less complicated than full-fledged sentences in regular text paragraphs.

Now, we will discuss briefly how to extract the  subject and the applied categorical attributes/constraints 
from a table. As mentioned before, we get a hint about the subject and the constraint of a table 
by parsing the title of the page from where the table is retrieved. Using this hint, we identify the
subject column in the table. To do so, we check whether any of the table's column headers matches with the subject retrieved from the page title. The match is 
considered true if any of the stemmed words (nouns) retrieved as subject from the page title 
matches with the stemmed column headers. Then, 
 the matched keyword is considered the subject for the training sample and the corresponding 
 column is identifies as subject column for that table. 
 
 There are few cases where no 
match was found with the table header,  as the subject obtained from the page title and the header 
of the subject column in the table do not use the same common noun for the subject. For example, 
in many cases, we found the table header of the subject column given simply as  `name'. 
In such cases, we use the retrieved subject from the page title as the subject for the training 
sample and the adjacent column to the  sortable column (ranking column) of the table is considered the
subject column. We filter the numeric attributes during our sample generation as presented in 
Line~\ref{alg:line11} of Algorithm~\ref{alg:1}.
To do so, we employ a dictionary of unites to recognize  numeric attributes, i.e., if the table header or cells contain 'lbs' or 'kg'.
 The presence of only  numeric content in a cell of a table column is considered as numeric attribute.
Although the structure of tables are well defined in Wiki\-pedia, the data is not free from ambiguity. 
For instance, in some tables, numerical 
data (e.g., age) are spelled-out, thus, are retrieved as non-numeric categorical values which 
leads to a false identification of numeric attribute as categorical attribute in our training data.
Here, more sophisticated extraction methods~\cite{DBLP:journals/pvldb/LimayeSC10, DBLP:conf/semweb/BhagavatulaND15} could be used in Algorithm~\ref{alg:1} for metadata extraction, but this is orthogonal to our hypothesis.
It should be emphasized, however, that despite this restriction in obtaining training data, the 
learned model can in fact predict interesting categorical attributes of numeric type.

{\bf Potential Limitations:} 
Not surprisingly, in various cases, the absence of a table in Wiki\-pedia might be due to the 
limited manpower and not due to general disinterest in that table. 
In fact, we found cases of tables missing in Wiki\-pedia where human evaluators in our 
user study unanimously state that they are interesting and, following our hypothesis, should exist. 
For instance, the list of the gold medalists in Olympic history, grouped by the type of sport, was 
not present in Wiki\-pedia at the time of harnessing training data, 
but marked interesting by the majority of voters in our user study. In fact, at the time of writing 
this paper, several such lists of gold medalists were added to Wikipedia. Due to this characteristics, 
the accuracy of the samples extracted by Algorithm~\ref{alg:1} suffers from false positive data 
and reaches only 68.9\% overall accuracy according to user assessments.
However, even though few training samples were apparently misleading, our classifier was 
able to correctly classify the task according to the evaluators' judgments.
Supported by such exemplary evidence, and the overall performance shown in the 
experimental evaluation, we believe the hypothesis is reasonable to generate labeled training 
data for our learning task, {\em as important tables are created supposedly before people spend effort 
in creating less important ones}.


\section{Learning The Classification Model}
 \label{sec:learning}
 
The first step toward creating the classification model consists of mapping 
 the retrieved training samples to a feature space that captures the characteristics of the samples. 
 As discussed earlier in Section~\ref{sec:intro}, {\it interestingness is not a fixed concept.}
It is rather a meta concept capturing various separate concepts like 
{\it coverage}, {\it reliability}, {\it peculiarity}, {\it diversity}, and {\it utility} of data. Here, we 
propose three  statistical measures, coined {\it P-Diversity}, 
{\it P-Peculiarity}, and {\it Max-Info-Gap} and we discuss four existing statistical measures,  
{\it Entropy}, {\it Max-Coverage}, {\it Unalikeability}, and {\it Peculiarity}. The features are first extracted from the training samples, then the $\nu$-SVM approach is used to learn the classifier.

Before we dive into the concrete definitions of the individual measures,
let us have a look at Table~\ref{tab:sample_measures} in order to understand better on which data 
these measures are actually executed. In the heading of the table, we see values and their frequencies 
(i.e., the set $V_{a_c}$) for three made-up exemplary tables for a categorical attribute `country'. 
We see that the different measures
vary quite strongly, relative to each other, but also compared to the same measure for different table 
characteristics. For instance, for the first table on the left side, Entropy ($\hat{H}$) is quite low compared to the 
almost uniform (random) distribution of frequencies in the table in the middle which has, thus, a high Entropy.
This is the key point---individual measures highlight different aspects of the data, for instance
the degree of randomness or the degree of dominance of categorical values.

\subsection{Existing Features for Categorical Attributes}
 \label{sec:features}
 
There are several probability-based objective measures proposed in literature, specifically for mining 
association or classification rules, capturing the generality or reliability of such rules. 
One of the most prevalent measures is {\em Entropy}, mainly used for mining attribute-value pairs. 
Statistical measures that are capturing diversity of categorical attributes are, on the other hand, less 
prominently investigated~\cite{DBLP:journals/isci/Domingo-FerrerS08}. 
Below, we briefly review four traditional measures and how to employ them 
as features to learn the classifier. 

\def\nameentropy{$\hat{H}$}
\def\namemaxinfogap{${mIg}$}
\def\namemaxcoverage{$mCov$}
\def\nameunalikeability{${U}$}
\def\namepeculiarity{${D}$}
\def\namepdiversity{$\hat{pVar}$}
\def\nameppeculiarity{$\hat{pPec}$}

 \begin{table}[!t]
\caption{Sample Data and Corresponding Measures}
\label{tab:sample_measures}

 {
\small
\begin{tabular}{|p{2.1cm}|l|l|l|} \hline 
& {\bf Example 1} & {\bf Example 2} & {\bf Example 3}  \\  
&USA(12) & USA(2) & USA(12)  \\  
&Spain(8) & Germany(2) & Germany(2)  \\  
&Germany(2) & China(2) & China(2)  \\  
&China(2) & Australia(2) & Australia(2)  \\  
&Australia(2) & France(2) & France(2)  \\  
&France(2) & Switzerland(1) & Switzerland(1)  \\  
&  &   Russia(1) & Russia(1)  \\ \hline 
\nameentropy & 0.44 & 0.77 & 0.48\\ \hline 
\namemaxcoverage & 0.43 & 0.17 & 0.55\\ \hline 
\namemaxinfogap & 0.75 & 0.28 & 0.8\\ \hline 
\nameunalikeability & 0.71 & 0.85 & 0.67\\ \hline 
\namepeculiarity & 0.74 & 0.92 & 0.7\\ \hline 
\nameppeculiarity & 0.58 & 0.33 & 0.69\\ \hline 
\namepdiversity & 0.36 & 0.66 & 0.49\\ \hline 
\end{tabular}

}
\end{table}

\begin{figure*}[t]
	\subfigure[Interesting samples]{ \label{fig:fstudy4} \includegraphics[width= 28mm, angle= 270]{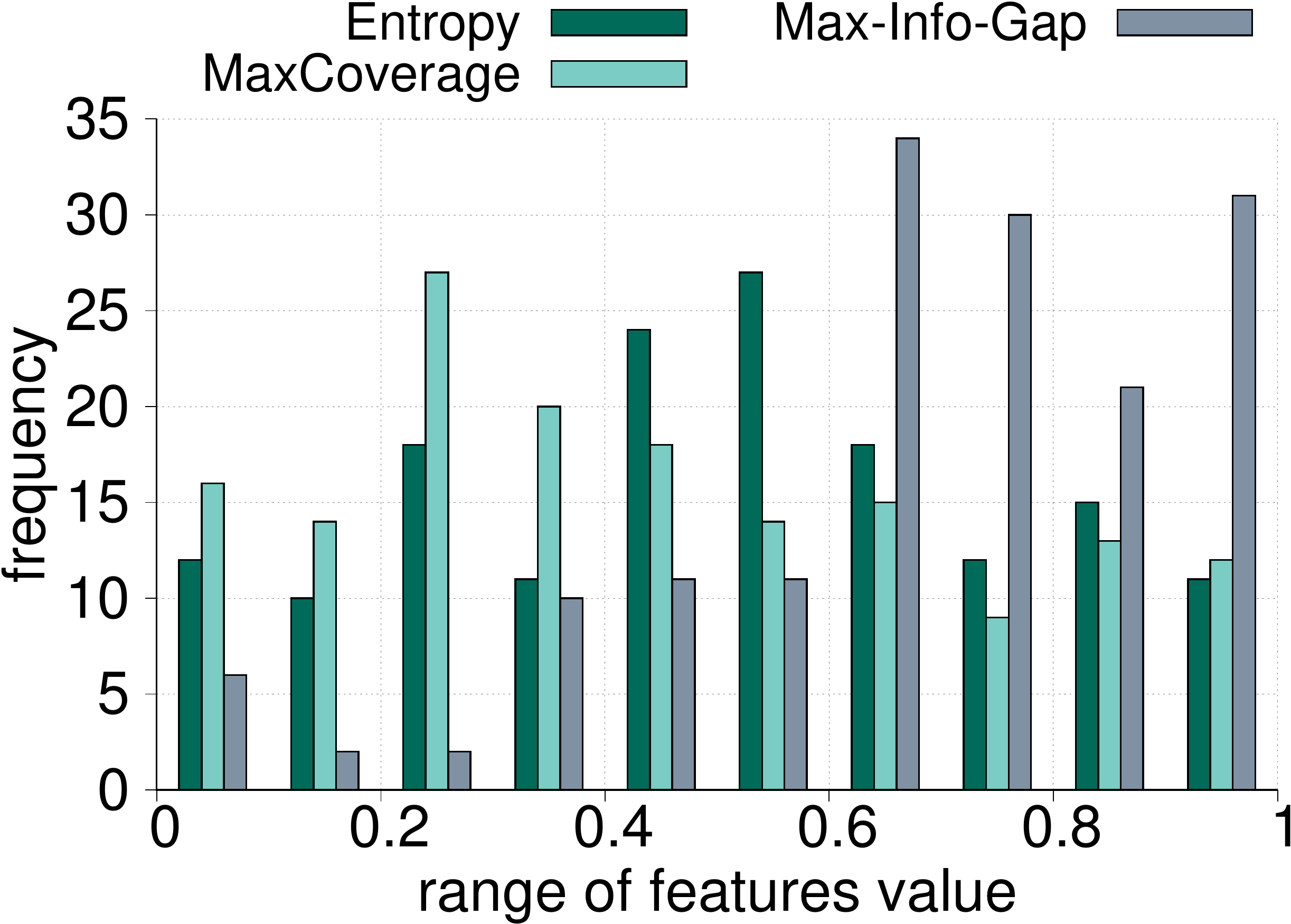}}  
       \subfigure[Non-interesting samples ]{\label{fig:fstudy2} \includegraphics[width= 28mm, angle= 270]{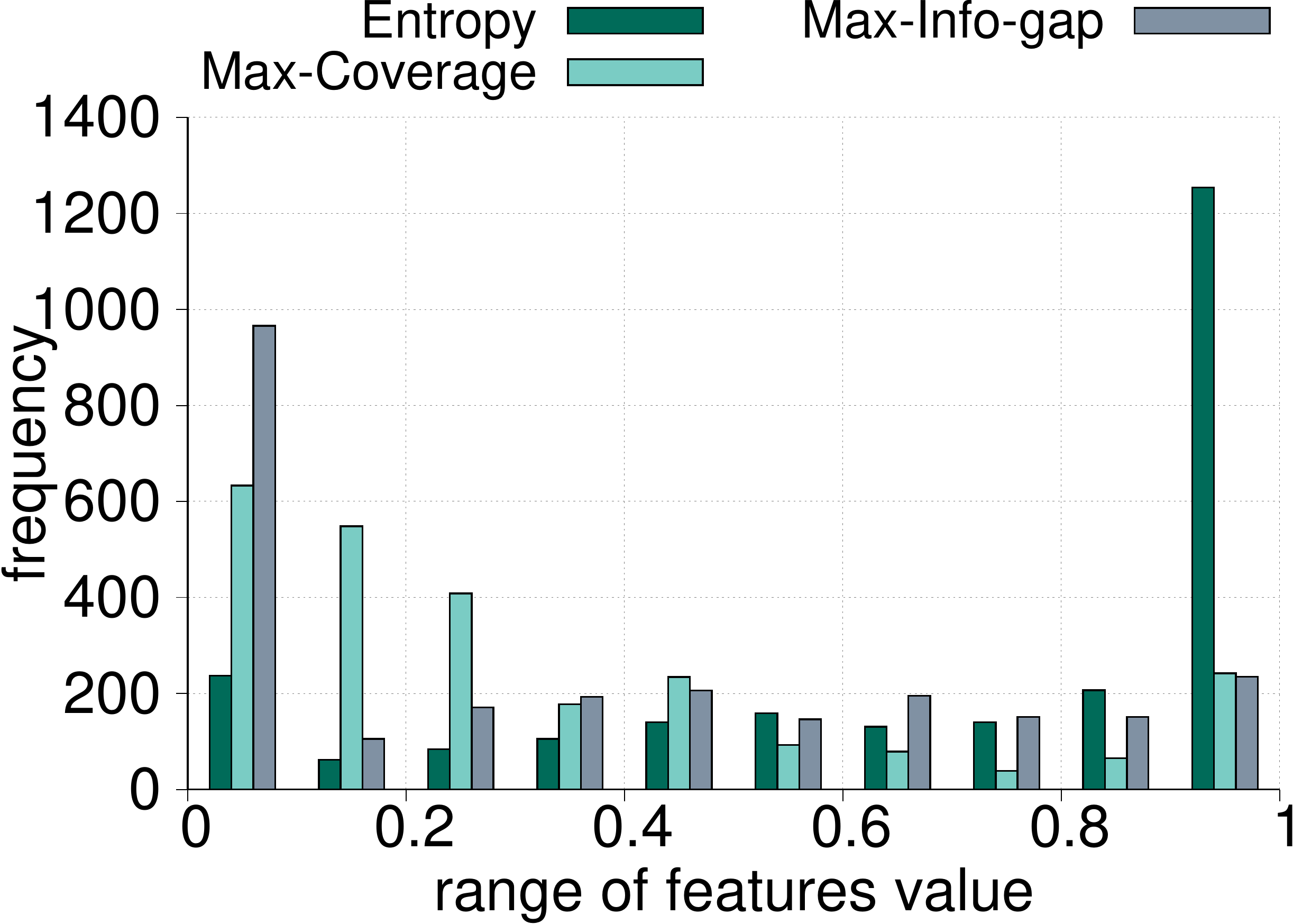}}     
       \subfigure[Interesting samples]{\label{fig:fstudy3} \includegraphics[width= 28mm, angle= 270]{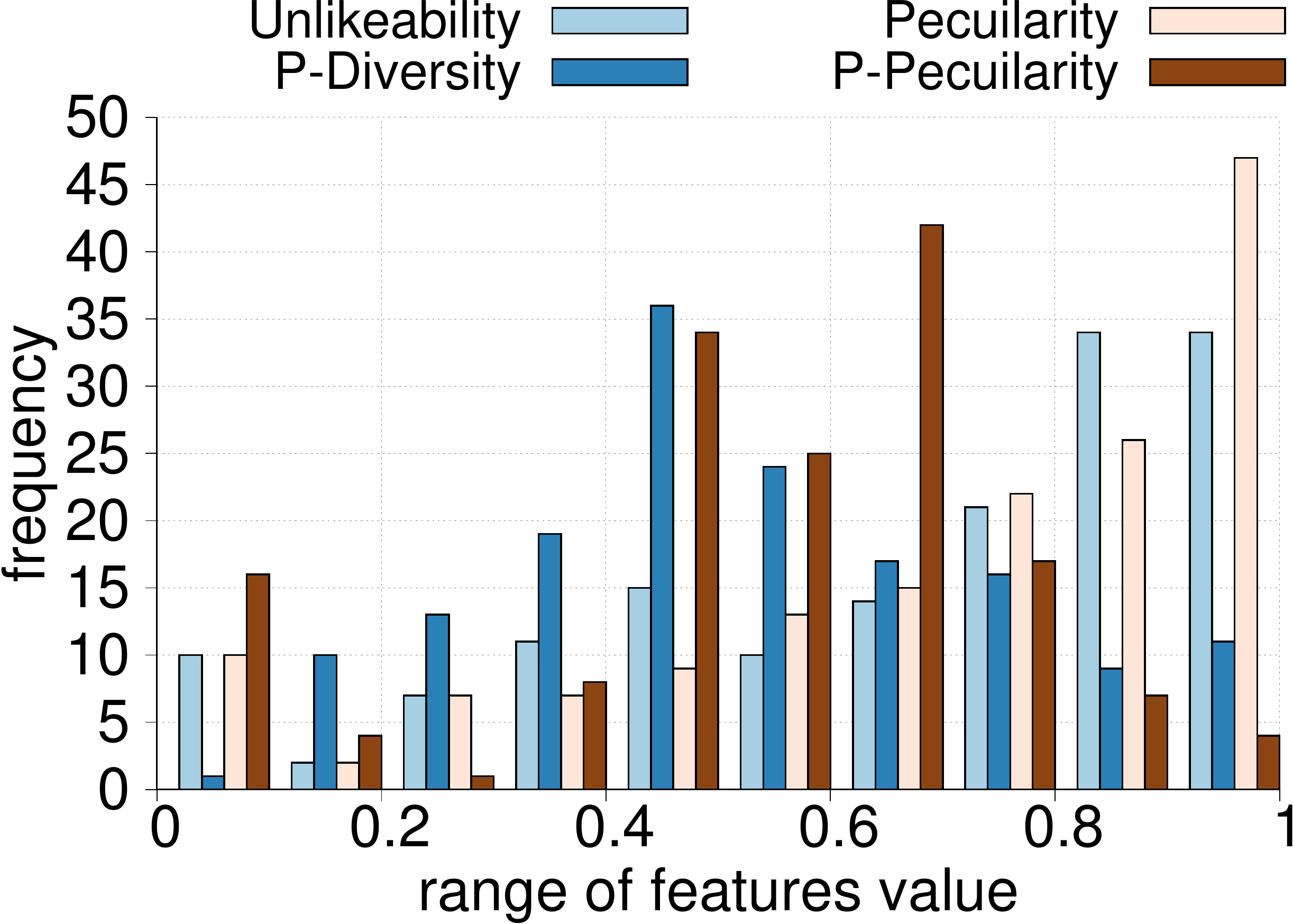}}        
      \subfigure[Non-interesting samples ]{\label{fig:fstudy1} \includegraphics[width= 28mm, angle= 270]{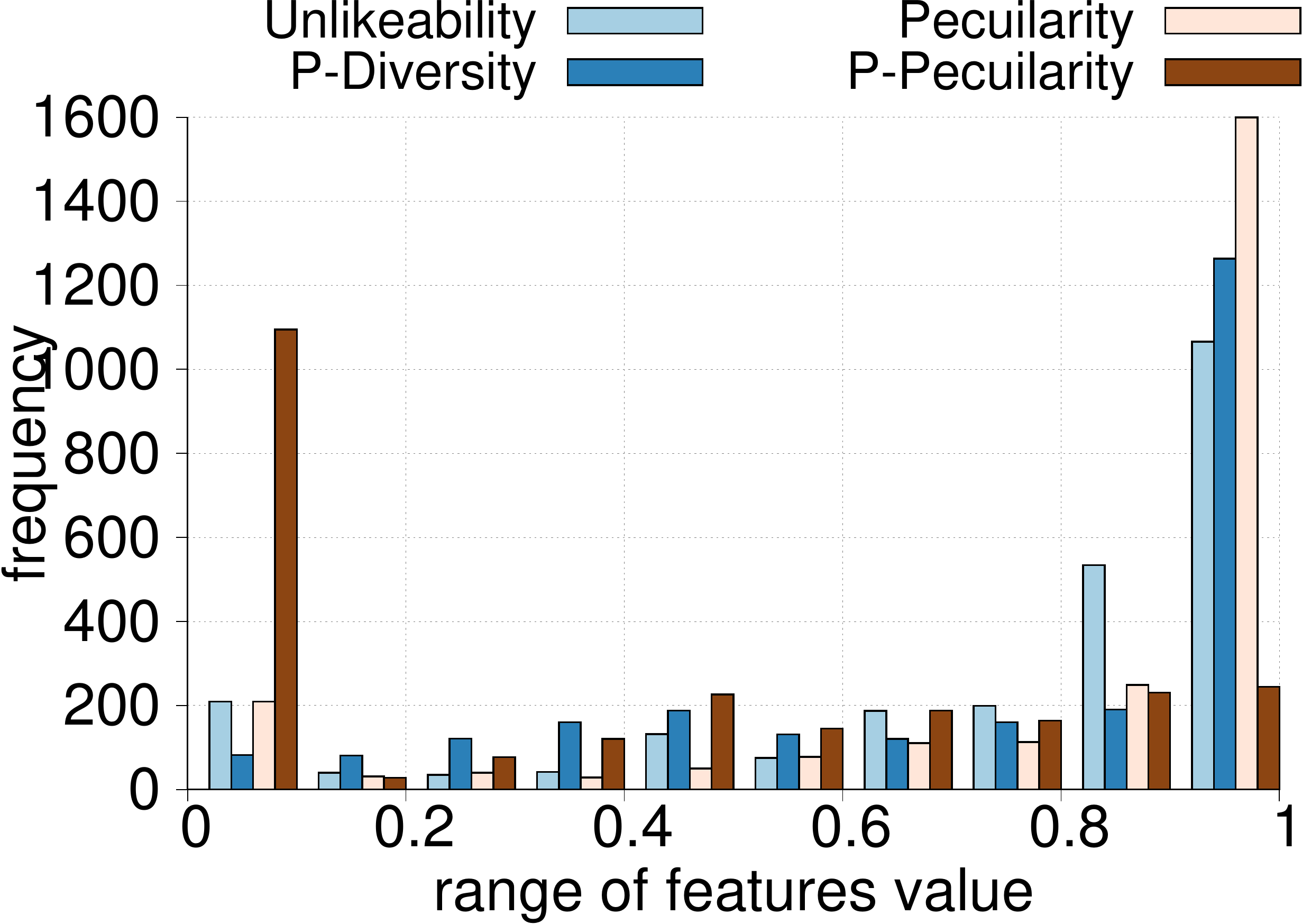}}       
       \caption{Distribution of features value}
    \label{fig:fstudy}
\end{figure*}  

{\bf \textsf{Shannon Entropy}}: 
Entropy reflects the degree of uncertainty in the 
information, described by a discrete random variable. In this work, a categorical attribute 
$a_c$ is treated as a random variable where $\mathcal{V}_{a_c}$ is 
the set of possible values that $a_c$ can hold. Shannon entropy for $a_c$ is calculated 
by $H(a_c) = -\sum_{x\in \mathcal{V}_{a_c}} P(x)\log_2P(x)$,
with $P(x) = count(x)/ |\mathcal{T}|$,  where $|\mathcal{T}|$ is the size of the table and $count(x)$ is the frequency of value $x \in \mathcal{V}_{a_c}$.
We use the normalized entropy, given by 
$$\hat{H}(a_c)=\frac{-\sum_{x\in \mathcal{V}_{a_c} } P(x)\log_2P(x)}{\log_2|\mathcal{T}|}
$$
How can we relate entropy to the interestingness of categorical attributes?
Intuitively, a piece of information is considered interesting when the randomness of the information
content is neither extremely high (i.e., $\hat{H}(a_c)=1$) nor extremely low 
(i.e., $\hat{H}(a_c)=0$). This interpretation of entropy in our context is 
reflected in Figure~\ref{fig:fstudy}\subref{fig:fstudy4} and~\ref{fig:fstudy}\subref{fig:fstudy2}. It shows that 69\% of the seemingly interesting
categories have an entropy value in $[0.2, 0.8]$ and 71\% of the non-interesting 
categories have an entropy value in $[0,0.2]$ or $[0.8,1]$. 

{\bf \textsf{Max-Coverage}}: Coverage is a commonly used measure in itemset mining. It captures
the comprehensiveness of a pattern. In this paper, we use maximum coverage of a categorical 
value as one of the features. It is denoted as $mCov(a_c)$, and is calculated as 
$$mCov(a_c) = max_{x\in \mathcal{V}_{a_c}} \{P(x)\} $$ 
$mCov$ captures dominance of a categorical value. If all the entities in the table have the same value for
$a_c$, then $mCov(a_c)=1$. Such an extreme case is definitely not an interesting one. On the other hand, 
$mCov(a_c) \rightarrow 0$  when too many categorical values are associated with $a_c$ and 
each entity holds a different categorical value. This scenario is also not an 
interesting one. Intuitively, a mid range in [0,1] might represent $mCov$-value for a interesting category.
According to the distribution of the measure $mCov(a_c)$ in our training samples,
presented in Figure~\ref{fig:fstudy}\subref{fig:fstudy4} and~\ref{fig:fstudy}\subref{fig:fstudy2}, 66\% of 
interesting categories have a \textsf{Max-Coverage} in $[0.2, 0.8]$ and  47\% of the non-interesting categories 
have a $mCov$-value in $[0, 0.2]$. 

It should be mentioned here that for large entity list, both cases, a skewed distribution
with  high $mCov$-value and a uniform distribution with lower $mCov$-value are
suitable for further categorization of entities. But these ranges of $mCov$-value identify 
the categorical attributes as not interesting for categorization. 

{\bf \textsf{Unalikeability}}: Variance is a common measure for describing the {\it degree of diversity} 
present in the data. For categorical data, Kader and Perry~\cite{journals/Gary07} 
discuss a variation coefficient, called Unalikeability, denoted as $U(X)$ for  a 
random variable $X$. Instead of measuring how much a observation of random variable differs, it essentially 
captures how often observations of a random variable differ from one another.
For a categorical attribute $a_c$, it is calculated as 
$$U(a_c) = 1- \sum_{x\in \mathcal{V}_{a_c} } P(x)^2$$
$U(a_c)=0$ shows that all the entities have the same value for $a_c$ and, thus, categorical attribute $a_c$ 
for that entity class becomes less interesting. Similarly, $a_c$ becomes non-interesting while 
$U(a_c) \rightarrow 1$. It signifies that $a_c$ is too diverse to choose an attribute value pair for further
refinement of entity list. Figure~\ref{fig:fstudy}\subref{fig:fstudy1} reflects this characteristic,  83\% of 
non-interesting categories hold Unalikeability values in $[0,0.2]$ or $[0.8,1]$.

{\bf \textsf{Pecularity}}: Simpson's index is the most commonly used diversity measure for categorical 
attributes and it is also referred as peculiarity measure in the context of mathematical ecology as discussed 
in Kader and Pielou~\cite{Pielou69}. It is defined by the probability that a randomly chosen categorical 
value has not been seen previously, denoted as $D(X)$ for random variable $X$. For a categorical 
attribute $a_c$, it is calculated as 
$$D(a_c) = 1- \sum_{x\in \mathcal{V}_{a_c} } \frac{count(x)(count(x)-1)}{|\mathcal{T}|(|\mathcal{T}|-1)}$$
$D(a_c)$ also shows the same characteristic as $U(a_c)$ with respect to understanding
the interestingness of categorical attributes. Figure~\ref{fig:fstudy}\subref{fig:fstudy1} reflects this 
characteristic of Peculiarity. Around 74\% of non-interesting categories hold feature values within the range of $[0,0.2]$ or $[0.8,1]$. Both diversity measures show that an ideal interesting category is more prone to have a diversity-value 
in mid-range of [0,1], near to 0.5. 
 
 \subsection{Novel Features for Categorical Attributes}
 \label{sec:novel}
 
The existing measures discussed above are commonly used as impurity measure in classification methods, 
such as decision trees, and are able to capture the distribution of categorical values. But 
capturing only the distribution or uncertainty in information content of categorical values for an entity class 
is not enough to understand whether that distribution would be interesting enough for further categorization 
of the entities, or not. The problem we are addressing in this paper calls for a more fine-tuned understanding of the
data distribution. In the following, 
we propose three probability-based statistical measures: (i) \textsf{Max-Info-Gap}, (ii) \textsf{P-Diversity}, 
and (iii) \textsf{P-Peculiarity}. We also discuss how these measures provide more distinctive information to 
understand which distribution of categorical values would be interesting in our context, compared to the above existing 
statistical measures.
 
{\bf  \textsf{Max-Info-Gap}}: 
Consider a specific value $x$, say `China', of a categorical attribute $a_c$ in a table.
If $x$ is very frequent, the information contained in that column of the categorical value is low,
with the extreme case of one unique value for all entities with respect to $a_c$. This extreme case could in 
fact indicate that the table was explicitly created for entities that hold this one specific value.
Following information theory, the maximum amount of information that a specific categorical
value can hold is $ - \log_2 \frac{1}{|\mathcal{T}|}$ for a table $\mathcal{T}$ with $|\mathcal{T}|$ rows;
basically when it is describing only one entity in the table.
Now, the idea behind Max-Info-Gap is to quantify the maximum difference between the information expressed
by one specific categorical value within $V_{a_c}$ and the maximum information content a categorical value 
can hold hypothetically for that table,
given by $$max_{x\in{V_{a_c}}}\left\{(-\log_2 {|\mathcal{T}|^{-1}})-(-\log_2 P(x))\right\}$$

The maximum difference occurs due to the categorical value having maximum coverage.
Compared to Max-Coverage, that is concerned with solely maximizing $P(x)$,
Max-Info-Gap specifically incorporates the size of the table. It signifies the dominance of a categorical 
value compare to others. With the existing notion of Max-Coverage, we can define Max-Info-Gap as follows.
\begin{definition}
Max-Info-Gap is the maximum information gap between the maximum information that a 
categorical value can hold for the categorical attribute and the actual information it is
 holding. It is denoted as $mIg(a_c)$ for categorical attribute $a_c$, and is calculated as follows:
$$mIg(a_c) = 1- \frac{\log_2 mCov(a_c)}{\log_2 |\mathcal{T}|^{-1}}$$
\label{def2}
\end {definition}
The values of $mIg(a_c)$ fall by definition into $[0, 1]$. Full diversity in the values of $a_c$ renders 
$mIg(a_c)=0$, clearly not a interesting scenario for further categorizing entities.
For a fixed table size, as the dominance of one specific value increases, $mIg(a_c)$ also increases. 
In general, we can say that a skewed distribution of categorical values hold higher $mIg$ measure than 
the uniform distribution for fixed table size.  In extreme case, when all entities hold the same categorical 
value then $mIg(a_c) = 1$ as $mCov(a_c) =1$. For example, if we compare the values of
$mIg$ in Table~\ref{tab:sample_measures}, we find that the $mIg$-value of the Example 1 and 
Example 3 are higher than for Example 2 as the distribution of categorical values in Example 1 and 
Example 3 are skewed. We can also see that  $mIg$-value of Example 3 is slightly higher than for 
Example 1, as the  dominance of `USA' is higher in Example 3, considering that the table length of both examples is comparable. 

\begin{figure}[t]
    \centering    
   \includegraphics[height=0.7\columnwidth, angle=270]{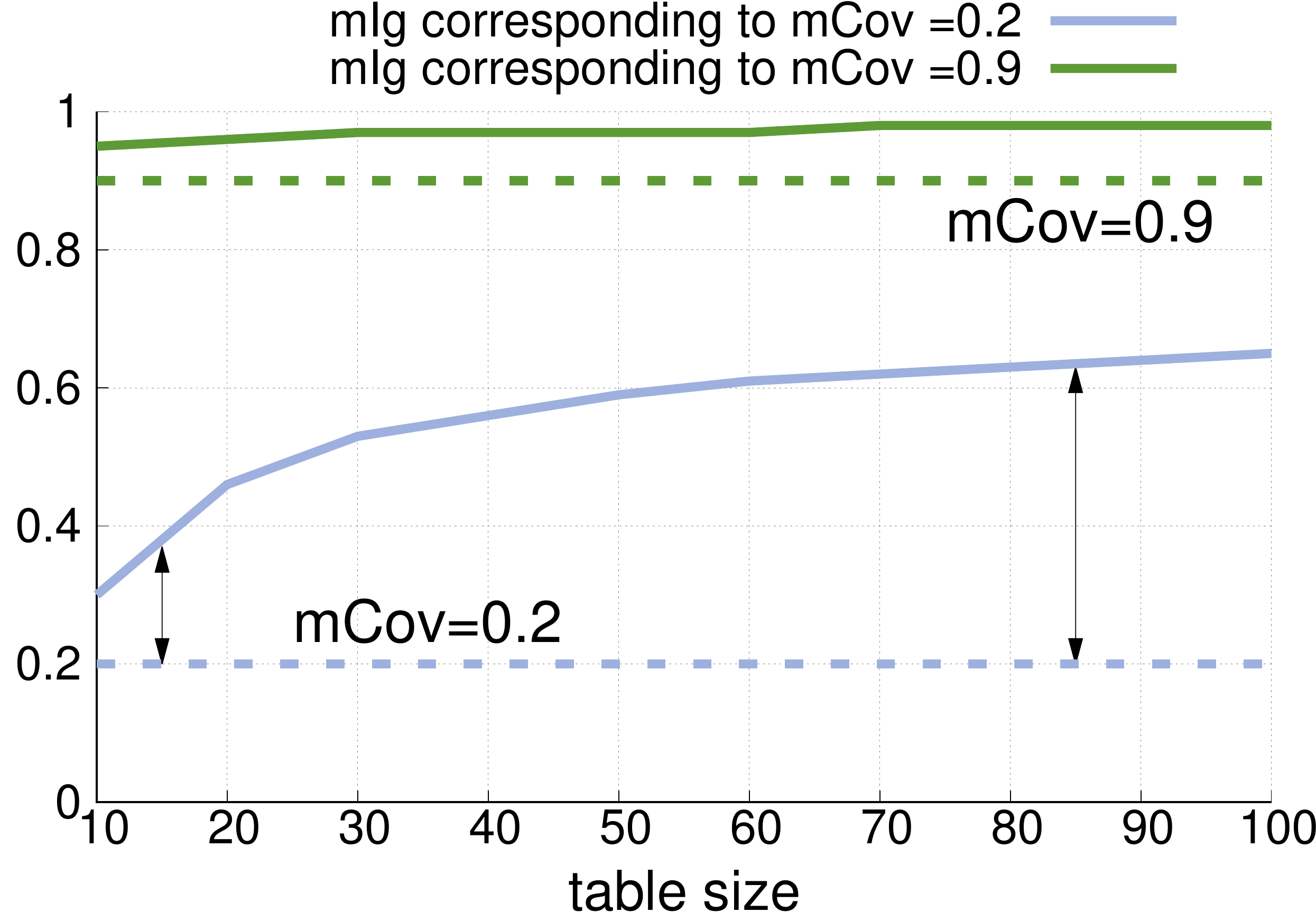}
     \caption{Comparison between \textsf{Max-Info-Gap} and \textsf{Max-Coverage} with varying table size.}
    \label{fig:mIg}
\end{figure}

It is important to discuss here how significantly $mIg$  differs from $mCov$, as both of these measures 
quantify the dominance of categorical value in a table. Both measures, $mIg$ and $mCov$ hold a 
value towards $1$ if one categorical value is very dominant. But  $mIg$ considers the table length to 
reward the coverage value as table length increases. In Figure~\ref{fig:mIg}, we present how the measure 
$mIg$ rewards coverage value with varying table length. We can see from the Figure~\ref{fig:mIg}
that for higher $mCov = 0.9$, $mIg$-value does not differ much which clearly signifies existence of one 
very dominating categorical value. On the other hand, with low coverage value, representing
a non-skewed distribution of categorical values, significantly differs from $mCov$-value as table length 
increases. In Figure~\ref{fig:mIg}, we can see that with a fixed low $mCov = 0.2$, a small table of length 10 
holds $mIg = 0.3$ where a larger table with 100 entities reach $mIg = 0.63$, significantly higher to indicate a 
further categorization of entity based on that categorical value. The $mCov$ measures cannot capture this 
insight from data. 

\begin{table}[h]
\small
\begin{tabular}{|p{0.124\columnwidth}p{0.124\columnwidth}p{0.124\columnwidth}p{0.124\columnwidth}p{0.124\columnwidth}p{0.124\columnwidth}|} \hline
\multicolumn{6}{|l|}{{\bf Example} 4}\\ \hline
USA & Spain& Germany & China & Australia &France   \\  
(60) & (50) & (45) & (60) & (40) & (60) \\  \hline 
\end{tabular}
\end{table}

From Example 2  and Example 4, we can see that 
Example 4 is more suitable for further categorization compared to Example 2. Both examples have $mCov = 0.16$ but 
$mIg$-value for Example 4 is $0.74$, emphasizing a possible interesting further categorization; whereas the $mIg$ value of 
Example 2 is very low (cf., Table~\ref{tab:sample_measures}). 
In line with the discussion, Figure~\ref{fig:fstudy}\subref{fig:fstudy4} and~\ref{fig:fstudy}\subref{fig:fstudy2} shows that 19\% interesting and 
24\% non-interesting samples have $mCov \leq 0.1$ in our training data whereas 3\%  of the interesting and 39\% of the 
non-interesting samples have $mIg \leq 0.1$. This means, Max-Info-Gap ($mIg$) can better distinguish  interesting categorical attributes from 
 non-interesting ones.\\

{\bf \textsf{P-Diversity}}: To understand the deviation of the distribution of categorical values from
a predefined reference distribution, we propose the P-Diversity measure.
In contrast to the concept of Unalikeability $U(a_c)$, it aims at describing how often
the observation of a random variable varies with respect to an established reference frequency. 
This reference frequency is defined based on how we define the interesting categorical attribute. 
For a categorical attribute $a_c$, if all entities have the identical value then there is no diversity 
among the observations (in such case, $U(a_c)=0$). In this scenario, imposing 
a constraint on $a_c$ cannot create a new, refined table. Hence, $a_c$ is not  interesting. 
So, the categorical attribute needs to have at least two values, introducing the possibility of 
having refined tables that are created by putting a constraint over values of
$a_c$. In an ideal case, these two categorical values will be equally distributed over 
the entity list which defines the reference frequency $0.5$.
It represents the minimum diversity for a categorical attribute to become interesting. 
Relative to this reference frequency, we define the measure P-Diversity of a categorical attribute as follows.
\begin{definition} P-Diversity is the square root of the sum of squares of the differences between the actual coverage of a categorical value to the reference coverage value $0.5$. It is denoted as  $pVar(a_c)$ 
for $a_c$, and is calculated as 
$$pVar(a_c) = \sqrt{\sum\limits_{x\in \mathcal{V}_{a_c}} (P(x)-0.5)^2}$$ 
\label{def3}
\end{definition}
Note that the 
maximum $pVar(a_c)$ is equal to $(1-0.5|\mathcal{T}|)/\sqrt{n}$, which occurs whenever the categorical attribute 
holds different values for each entity in the table. 
The normalized P-Diversity is, thus, given by
 $$\hat{pVar}(a_c) = \frac{\sqrt{\sum\limits_{x\in \mathcal{V}_{a_c}} (P(x)-0.5)^2)}}{(1- 0.5|\mathcal{T}|)/\sqrt{n}}$$ 

Here, $\hat{pVar}(a_c)=1$ if the categorical attribute holds different values for each of the entity. Therefore, 
$\hat{pVar} \rightarrow 1$ signifies the cases where further categorizations are not suitable. On the other 
hand, a $\hat{pVar} \rightarrow 0$ while the coverage of categorical values near 0.5, indicates the possibility 
of further categorization. The normalizing factor for \textsf{P-Diversity} also considers the size of the table 
and reward ${pVar}$ value as size increases. Hence, a uniform distribution with 
coverage value far from 0.5 might hold $\hat{pVar}$ value close to 0 if table size is large. Rewarding the 
table with larger size even for small coverage value is perfectly in line with the consideration for further 
categorization, in our context. For example, let us consider Example 2 and Example 4 where both lists have
 almost similar distribution of categorical values where coverage of each categorical value  is
less than $0.2$, significantly smaller than the reference coverage of $0.5$. But due to large table size, 
Example 4 holds $\hat{pVar} = 0.09$, close to $0$, indicating further categorization of the table while  
Example 2 holds $\hat{pVar} = 0.66$, indicating less importance of a further categorization of the entities. 
The measure Unalikeability cannot capture this characteristic and holds almost similar $U$ value 0.85 and 
0.83 respectively for Example 2 and Example 4, respectively,  indicating both are not suitable for further categorization of 
entity list. We can observe from the distribution of $\hat{pVar}$-value, shown in Figure~\ref{fig:fstudy}
\subref{fig:fstudy1} that 50\% of retrieved non-interesting training samples holds $\hat{pVar}$ value in range
 of [0.9, 1] and 72\% non-interesting training samples $\hat{pVar}$ value $\geq 0.5$.
 \begin{table}[h]
 \small
 \caption{Comparing $\hat{pVar}$-value with Unalikeability}
\label{tbl:U_Var}
\begin{tabular}{|c|p{3.7cm}|c|c|c|} \hline
 Table size&list & $\hat{pVar}$ & U & D \\ \hline
\multirow{ 2}{*}{100} &list1: \{ USA(80), Spain(20)\}& 0.09& 0.32 & 0.32 \\  
&list2: \{USA(60), Spain(40)\}&  0.03  & 0.48  & 0.48 \\   \hline  
\multirow{ 2}{*}{5} & list3: \{USA(4), Spain(1)\}& 0.63 &0.32 & 0.4 \\   
&list4: \{USA(3), Spain(2)\}&  0.21  & 0.48 & 0.6\\   \hline 
\end{tabular}
\end{table}
According to the definition of $\hat{pVar} $, a uniform distribution of categorical values is considered to be more 
useful for further categorization compared to a highly skewed distribution where $\hat{pVar} \rightarrow 1$. 
We can see in Table~\ref{tab:sample_measures} that Example 1 has lower $\hat{pVar}$ value compared 
to Example 3 because Example 1 is less skewed than Example 3 . 

What exactly is the difference between P-Diversity and existing diversity measures?
Let us consider the example in Table~\ref{tbl:U_Var}. In the first row, we represent two lists where list1 is more 
skewed than the list2. The second row holds list3 and list4 with an identical distribution of values as list1 and list2, 
respectively, but both lists are of smaller  size. As \textsf{Unalikeability} does not consider the list size, it cannot 
distinguish between list1 and list3 as distribution of both lists are same. In contrast, $\hat{pVar}(list1)$ is very 
close to 0, indicating the potential for further categorization of list1 whereas $\hat{pVar}(list3) = 0.63$, which tells that list3 is not interesting for any further categorization of its entities.
 On the other hand, the \textsf{Peculiarity} measure increases as the table size decreases. Hence, 
for a skewed distribution of values, a small table is considered more interesting than the larger one. 
For example, Table~\ref{tbl:U_Var} shows that $D(list3)= 0.4$ is closer to 0.5 than $D(list1) = 0.32$. As
mentioned earlier, $D$-value close to 0.5 characterize interesting attributes, and thus list3 is more suitable for 
categorization than list1, according to \textsf{Peculiarity}, which is clearly not the case.  
In this table, we can also see that as list size grows the difference between
$\hat{pVar}$-value for skewed and uniform distribution decrease as expected.

In the experimental study in Section~\ref{sec:experiments}, we will see 
that the learning model created by using P-Diversity performs better than the model that is considering  
existing diversity measures. \\

Next, {\bf \textsf{P-Peculiarity}} is proposed to capture the unexpectedness of categorical values that are distributed over the table. 
 There is no peculiarity in a categorical attribute if the categorical values are equally 
 distributed over the entities. Thus, this measure quantifies the skewness of data by finding the difference
of its distribution from uniform distribution. Unlike the $mIg$ measure, where the skewness of the most skewed categorical value is used to quantify the 
skewness, here, all categorical values is considered i.e., the actual probability distribution of the categorical 
attribute is considered.
 \begin{definition}
 P-peculiarity is the absolute difference between the actual probability distribution of the categorical 
 values and the uniform probability distribution.
 P-peculiarity is denoted as $pPec(a_c)$ for categorical attribute $a_c$, 
 and is given by
$$pPec(a_c) = \sum\limits_{x\in \mathcal{V}_{a_c}} |P(x) - \frac{1}{|\mathcal{V}_{a_c}|}|$$
 \label{def4}
 \end{definition}
 Again, we now investigate how this measure can be normalized. We observe that the maximum deviation from the 
 uniform distribution occurs when all categorical values,
 except  one, occur exactly once. The remaining one occurs for all 
other entities in the table. Hence,  P-Peculiarity normalized to $[0,1]$ by factor $max(pPec(a_c))$ is given by:

 $ \small{(| \mathcal{V}_{a_c}|-1)\left| \frac{1}{|\mathcal{T}|} - \frac{1}{ |\mathcal{V}_{a_c}|}\right| + \left|\frac{|\mathcal{T}|-|\mathcal{V}_{a_c}|+1}{|\mathcal{T}|} - \frac{1}{ |\mathcal{V}_{a_c}|}\right|}$. \\
 
 The normalized P-Peculiarity, $\hat{pPec}(a_c)$, is then simply given by
 
  $$\hat{pPec}(a_c) = \frac{\sum\limits_{x\in \mathcal{V}_{a_c}} |P(x) - \frac{1}{|\mathcal{V}_{a_c}|}|}{max(pPec(a_c))}$$\\
According to the formulation of $max(pPec(a_c))$, it is clear that $\hat{pPec}(a_c)=1$ 
 indicates that one specific  categorical value has almost full coverage over the entities. 
On the other hand, $\hat{pPec}(a_c) =0$ only when each categorical value has exactly same coverage. 
Both cases are considered to be not-interesting for further categorization of entity list. Our intuition is that 
$\hat{pPec}(a_c)$-value in mid range of [0,1] is considered to be interesting for further categorization of 
entity list. From Figure~\ref{fig:fstudy}\subref{fig:fstudy3} and Figure~\ref{fig:fstudy}\subref{fig:fstudy1}, we can observe this characteristic of distribution of $\hat{pPec}(a_c)$-value; 78\% of the interesting 
categories hold $\hat{pPec}$-values in $[0.2, 0.8]$ and 67\% non-interesting 
categories hold $\hat{pPec}$-values within $\leq 0.2$ and $\geq 0.8$. 

\begin{table}[h]
\small
\caption{Comparing $\hat{pVar}$-value with $mIg$}
\label{tbl:P_Var}
\begin{tabular}{|c|p{4.43cm}|c|c|} \hline
 Table size&list & $\hat{pPec}$ &  $mIg$\\ \hline
 100 &list1: \{ USA(90), Spain(10)\}& 0.82& 0.98\\  
10 & list2: \{USA(9), Spain(1)\}&  1.0  &   0.95 \\   \hline  
\end{tabular}
\end{table}

Similar to the $mIg$ measure, the normalizing factor used in P-Peculiarity also rewards the 
$\hat{pPec}$ value as table size increases. But unlike $mIg$, this measure incorporates the 
influence of the table size in skewed distributions. For example, Consider Table~\ref{tbl:P_Var}, we can see 
for list2, $\hat{pPec} =1$, clearly indicating not a interesting case for further categorization of entity list whereas 
$mIg= 0.95$ expressing an interesting case for further categorization  (recall that $mIg \rightarrow 1$ indicates
interesting scenarios). On the other hand, we can see, as the table size increases for  the same skewed distribution, 
the $\hat{pPec}$ value comes close to the mid range for list1, signifying an increasing interestingness  of a further categorization.

\subsection{Tailoring Support Vector Machines}
\label{sec:svm}
After having discussed possible features, we now describe how a classifier is created based on combinations of them.
We opted for applying the support vector machine (SVM) approach, a widely known and well understood concept. 
In the easiest case, traing data is balanced (roughly the same number of positve and negative samples) and linearly separable, which renders the application of simple Linear-SVMs possible.
 However, as we extracted the training data  from  Wikipedia it contains noise for the various reasons discussed earlier.
Hence, we need to employ a  {\it soft-margin classifier}, specifically, $\nu$-SVM~\cite{DBLP:journals/neco/ScholkopfSWB00} 
which can detect outliers while learning the classification model from the training data. In $\nu$-SVM, 
the parameter $\nu$ is tunable within $[0,1]$ and controls  the lower and upper bound on the number of 
misclassified samples that are allowed in the training phase of the classifier, i.e., the  training 
error. As $\nu$ increases the model becomes more biased and it is under-fitting the data. Moreover, 
we also notice that the statistical measures that we are going to use as feature space are  not linearly 
separable. Because for a statistical measure, multiple non-contiguous ranges of values can be associated 
with a particular class of samples. Hence, we use the popular Radial Basis Function (RBF) as kernel 
function~\cite{Scholkopf:2001:LKS:559923} that transforms our 
training data to higher dimensional space using a non-linear mapping. This `kernel' method allows classifying the
training data linearly in higher dimensional space. In the RBF kernel, a parameter $\gamma$ is used to control
the radius of influence of the support vectors. For example, a high value of $\gamma$ discards the influence of 
$\nu$, and cannot prevent overfitting. In Section~\ref{sec:experiments}, we will discuss how the optimal 
values for $\nu$ and $\gamma$ are calculated to train the classifier.

In this work, we extracted 16 times more 
negative samples than positive ones from Wikipedia tables using Algorithm~\ref{alg:1} (see details in Section~\ref{sec:experiments}). For such unbalance 
training data, the one-class SVM~\cite{DBLP:journals/neco/ScholkopfPSSW01} approach can be employed, where  the classifier is trained based on the samples
 from a single class (either positive or negative training samples). Another option could 
 be the use of sampling to obtain balanced training data.
 In the next Section~\ref{sec:experiments}, we present a comparative study on the performance of 
the classification models which are trained by either $\nu$-SVM on balanced samples created from original training data or one-class SVM on unbalanced original training data.

\subsection{Evaluation Methodology}
\label{sec:validation}
The learning model is validated in two different ways: {\it (i)} based on {\bf held-out test data} and {\it (ii)}  by means 
of a {\bf user study}.

For held-out test data, the samples extracted from Wikipedia tables using Algorithm~\ref{alg:1} 
are considered as ground truth to evaluate the performance of learned classification model. We use
accuracy as the performance metric. Class-specific accuracy  is the fraction of 
correctly predicted samples over all the test samples predicted to be that specific class according 
the classifier, which is also called precision for that specific class .
 
Our objective is to learn a classifier that can classify which categories are interesting to a user 
to categorize the entity list. Therefore, we also validate the classification model by means of a 
user study. As human-perceived interestingness is not a fixed concept,  
choices/preferences of users can differ. The human assessors are asked to classify 
each samples into one of three possible categories: 
{\bf\textsf{interesting}}, {\bf \textsf{non-interesting}}, and {\bf \textsf{not sure}}. 

Subsequently, we define ground truth depending on the agreement level of the human responses. 
The higher the agreement level (e.g., all agree on a label), the more ``obvious'' the task appears, 
and thus, it is presumably also easier for the classifier to correctly classify it. 
We see later that this is indeed the fact.

Consider a total of $y$ users that provide assessments of the test samples. 
Different agreement levels are considered based on majority voting: 
For the $x/y$ agreement level, where $x > y/2$, one of the three possible choices is considered 
as ground truth for a sample if at least $x$ users agree on that choice.  
The samples which are marked as {\it not sure} are excluded from the ground truth.  
For different agreement levels, class-specific accuracy/precision, recall, and F$_1$ measures
are used to validate the prediction of learning model against the ground truth, generated from 
the user study~\cite{DBLP:conf/icdm/WangY09}.  In order to quantify the reliability of user agreements, Fleiss' kappa~\cite{fleiss1971mns} is calculated.


\section{Experimental Evaluation}
\label{sec:experiments}

To obtained the training samples, we have used the English version of Wiki\-pedia dump file of 2016.
We have implemented Algorithm~\ref{alg:1} in Java 1.8. Experiments are executed on an Intel Core i7 CPU@3GHz
machine, with 16GB main memory. For the SVM classifier, the LIBSVM~\cite{CC01a}
library is used to create the learning models. 
The entire process of extracting the training samples from the 50.49GB (uncompressed)
Wikipedia dump took around 30 minutes, where most of the time was spent on actually cleaning-up the 
raw dump file before Algorithm~\ref{alg:1} was applied.
A total of 2045 ranking tables from Wikipedia pages entitled ``List of $\ldots$'' are extracted.
From these tables, based on Algorithm~\ref{alg:1}, $2519$ categorical attributes are labeled as ``non-interesting", 
are considered negative samples and $158$ categorical attributes are labeled as ``interesting", are considered 
as positive samples. 

For the training data, $75\%$ of positive and negative samples are randomly selected.
The remaining $25\%$ of samples from each class are considered as {\bf held out test data}, denoted
as \textbf{TestPos} and \textbf{TestNeg}, respectively positive and negative samples.
These two test datasets are merged into a set denoted as \textbf{Test}, containing  $669$ samples. 
We retrieved significantly fewer positive samples than negative samples. In order to create balanced  
training samples, we equally divided the negative samples into ten smaller chunks and then merged each 
of these chunks with the positive training samples, resulting in 10 sub-training files, each containing 306 training 
samples. The ratio of positive and negative samples in these sub-training files are $1\!:\!1.5$.

The labeled training data according to Algorithm~\ref{alg:1},
the 10 sub-training files, and the results of the user study are publicly
available online  on \url{http://dbis.informatik.uni-kl.de/catmining/} for repeatability and adoption.

\subsection{User-Study Setup}

As mentioned earlier in Section~\ref{sec:validation}, we set up a user study to validate the
trained classifier. To do so, $110$ randomly selected samples from Wiki\-pedia 
are presented to users. The samples are displayed to the users in form of ``$A (a_s):B$". In this 
format, `A' represents the title of the Wikipedia table, `$a_s$' represent the subject of the table, and
`B' represent a categorical attribute associated with the entity lists in the table. Users are asked 
to label the samples in three categories: {\it (i)} If a user is interested to categorize the entities in table 
`A' by using categorical attribute `B', then the sample is labeled as {\bf \textsf{interesting}},
 {\it (ii)} If a user thinks it is not interesting to categorize the entities in table `A' by using categorical attribute `B', then
the sample should be annotated with {\bf \textsf{non-interesting}},  {\it (iii)} a user can also label a sample {\bf 
\textsf{not sure}} in case the user can not decide any of the two options before.
 
Overall, each question is evaluated by $9$ human evaluators. Four evaluators are in fact enough 
to achieve a significance level\footnote{With 9 evaluators and 3 possible answers for each task, 
there are $3^9$ possible outcomes. Full agreement has, thus, a random chance of $3/3^9=0.00015$, 
$6/9$ agreement has random chance of $0.03$ for one-tail observation.} of $\alpha = 0.05$.
On average, an evaluator has marked $35.5\%$ questions with \textsf{interesting},
53.5\% questions with \textsf{non-interesting}, and $20\%$ with \textsf{not sure}.
As the user choices differ, we use Fleiss' kappa to understand the reliability of 
agreement. For each testing sample, {\it nine user choices} are taken for assessment. For
the nine evaluators, the calculated kappa value is $0.45$ and the 95\% 
confidence interval for Kappa has a range between $0.42$ and $0.48$ for the collected user data. 
 Moreover, this  range of values significantly differs 
from zero and  with $p$-value  $\approx 0$ (i.e., $\ll 0.05$), we can reject the {\it null hypothesis} that the 
agreement among users is achieved randomly. 

\subsection{Parameter Selection and Evaluation}

Due to the imbalanced size of the available training samples, discussed earlier,
it seems feasible to use a one-class SVM to learn a model {\it separately for each class}. 
Alternatively, we also have created the 10 balanced sub-training files from the original training sample,
as mentioned above.
Here, we evaluate the performance of classification models created by all feature combinations 
from $\mathcal{F}$, in total $2^{|\mathcal{F}|}-1$ combinations. The classification models 
are created for all possible feature combinations from each sub-training file. Modifying the 
parameter tuning in LIBSVM library, we implemented a grid search for $\nu$-SVM with 
$5$-fold cross-validation method to find the optimal parameter pair $(\nu, \gamma)$ for each 
sub-training file. Then, the classification model is learned with optimal $\nu$ and $\gamma$ value. 
According to the theoretical discussion in~\cite{Chen05atutorial}, for our training set, the solution 
of $\nu$-SVM is only feasible with $0\leq \nu \leq 0.77$. In fact, in line with the theoretical study, we found 
that the optimal $\nu$-value lies in $[0.41, 0.61]$ for different sub-training files with optimal 
$\gamma=0.0003$. For each feature combination, the average training time of 10 sub-training files 
is $13.813s$.

\subsection{Results Based on Held-Out Data}

\begin{figure}[!t]
    \centering    
   \includegraphics[height=0.7\columnwidth, angle= 270]{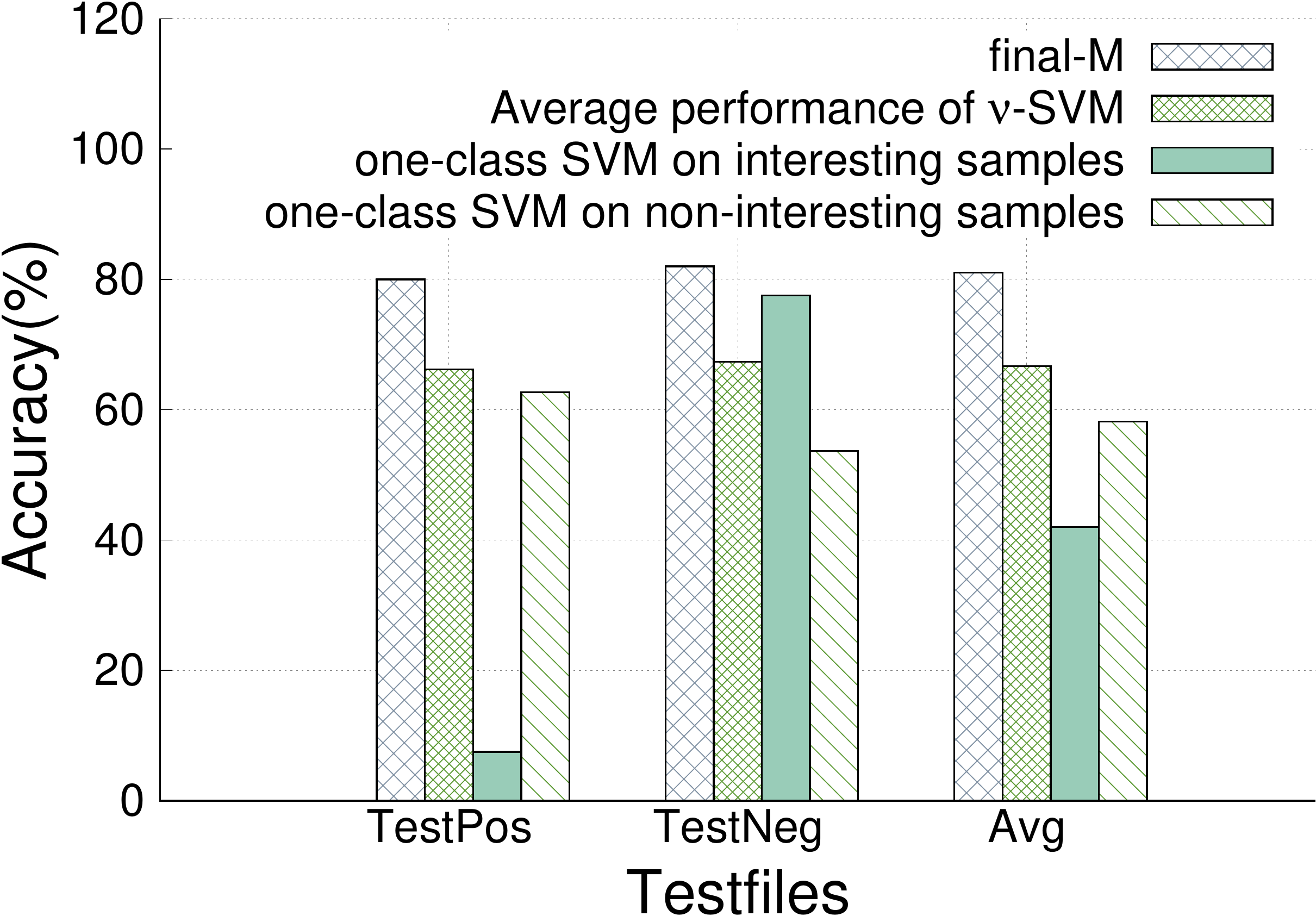}
     \caption{Comparison among different type of classification models}
    \label{fig:one-svm}
\end{figure} 
\begin{figure*}[!t]
    \centering    
     \subfigure[Recall] {\includegraphics[width= 33mm, angle= 270]{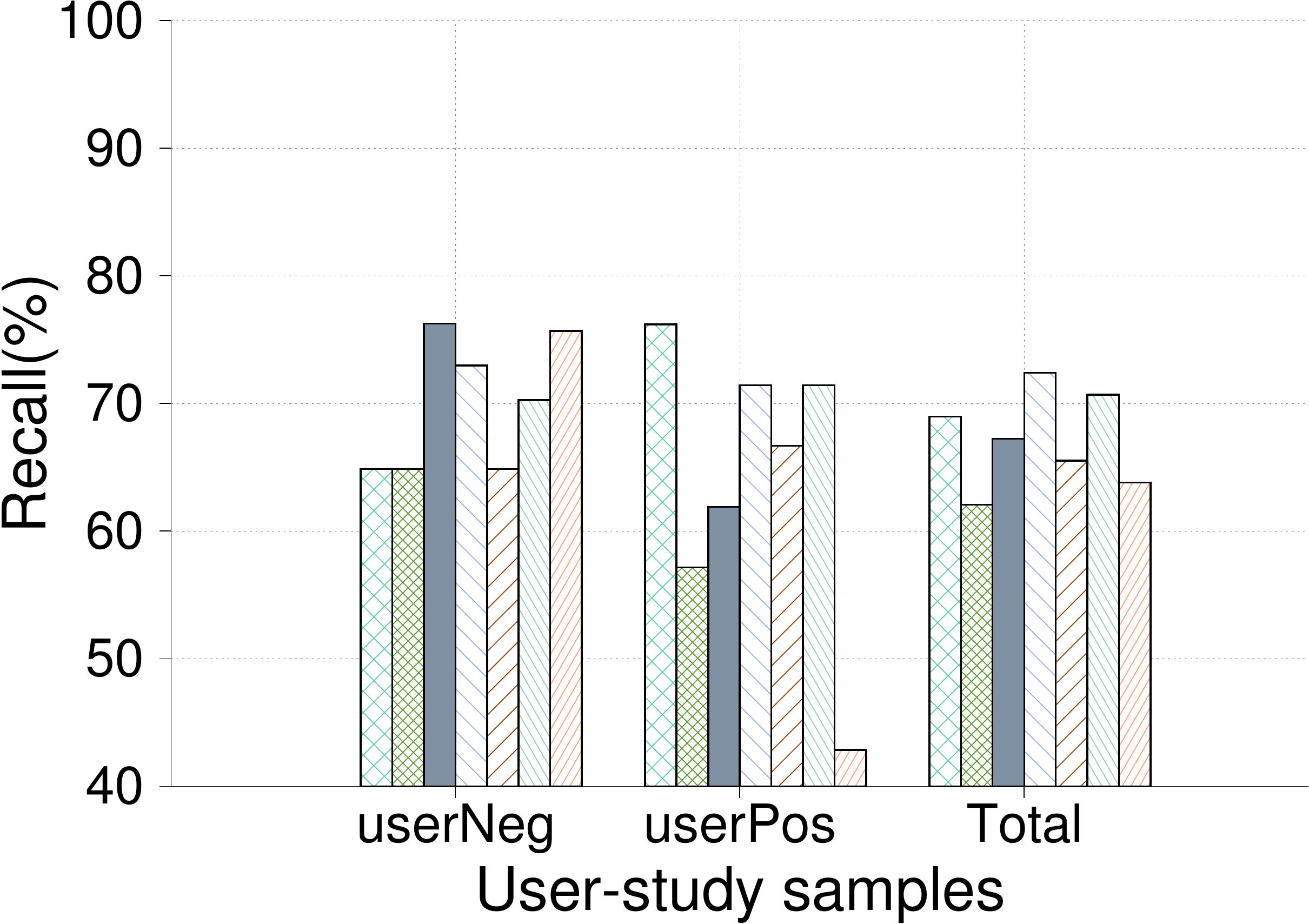}}
     \hspace{1pt}
       \subfigure[Precision] { \includegraphics[width= 33mm, angle= 270]{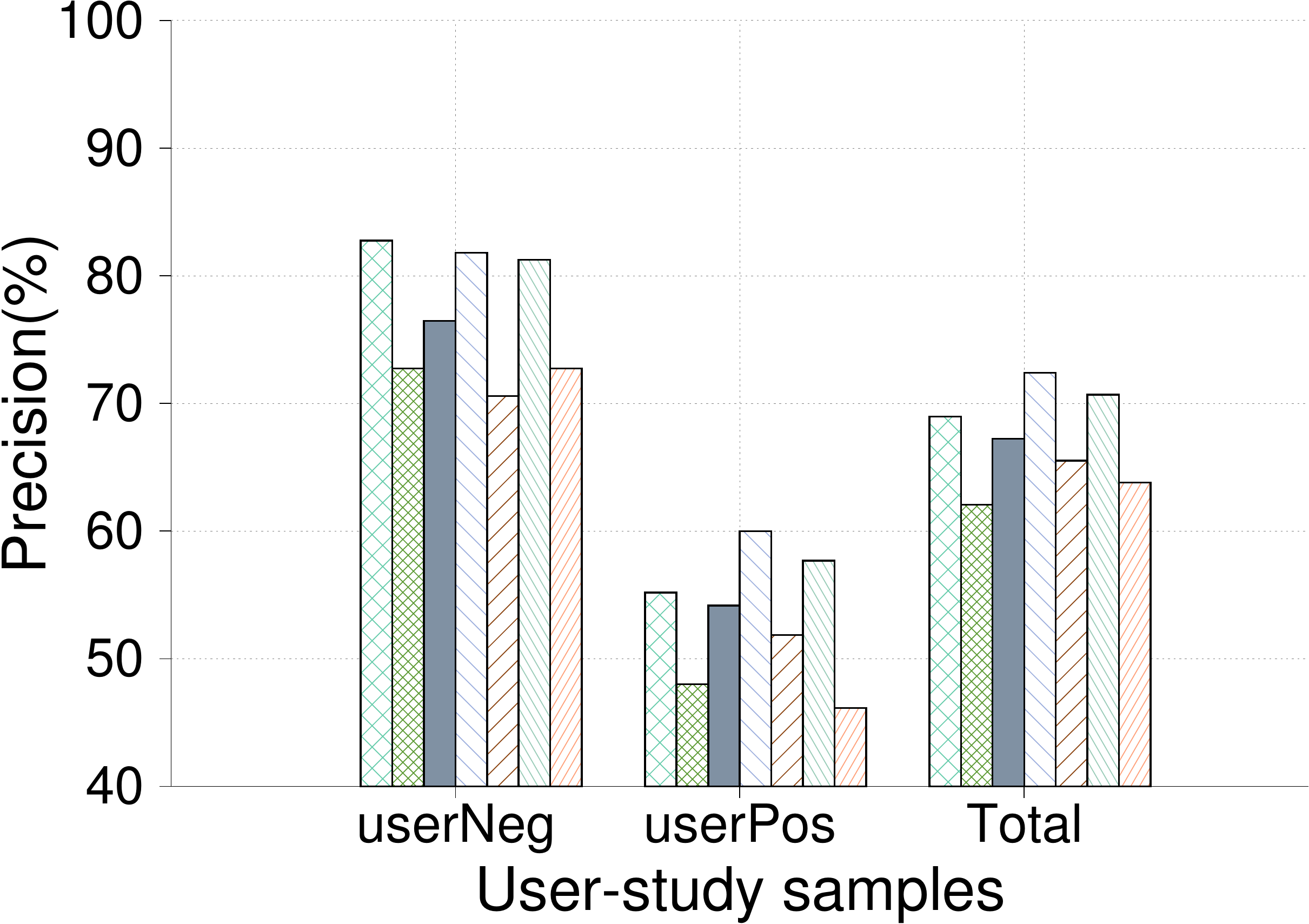}}
        \hspace{2pt}
         \subfigure[$F_1$-measure]{\includegraphics[width= 34mm, angle= 270]{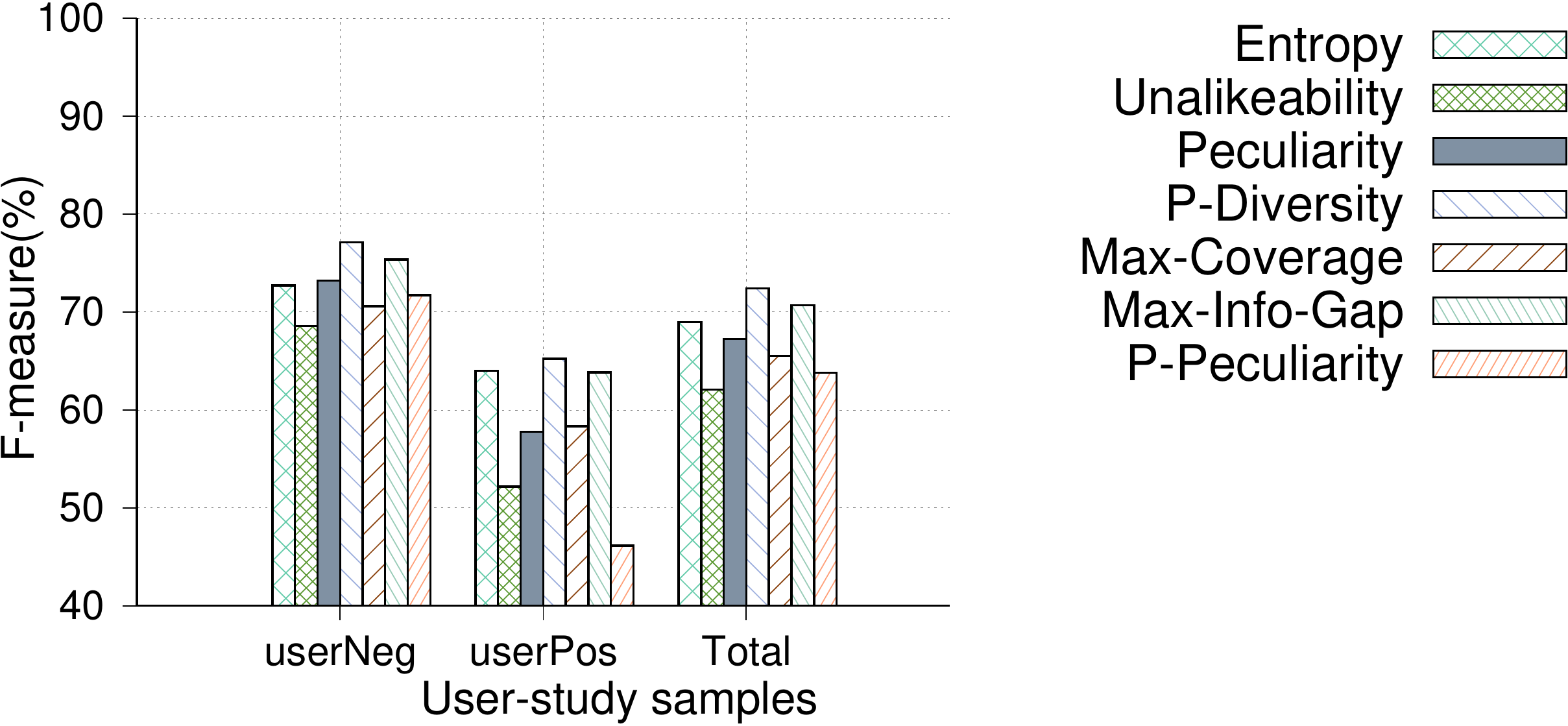}}
        \label{fig: recall}
       \caption{Comparison among classification models using single feature considering 6/9 agreement level as ground truth}
    \label{fig:featurecom}
\end{figure*} 

The performance of the classification models created on sub-training files is first
{\bf evaluated on held-out test data}. The held-out test data is created by random selecting 
of 25\% samples from complete samples. As complete samples retrieved from Wikipedia using
Algorithm~\ref{alg:1} is unbalanced, reflecting the same characteristic of original data, 
the held-out test data also contains 16 times more non-interesting samples than interesting ones.
Therefore, a high overall accuracy on \textbf{Test} data does not imply that the class-specific 
accuracy is also high, i.e., the model performs well for both \textbf{TestPos} and \textbf{TestNeg} 
data separately. Hence, rather considering the overall accuracy of the classification models on 
\textbf{Test}, we consider class-specific accuracy i.e., precision of \textbf{TestPos} and \textbf{TestNeg} 
separately to evaluate the classification model. For each feature combination $f \in 2^\mathcal{F}$,
we train 10 classifiers based on the 10 sub-training files and choose the ones, called  
\textbf{$\textsf{best}_f$}, that have minimum classification error on {\it both} the 
\textbf{TestPos} and \textbf{TestNeg}. Finally, the classification model which performs best 
among all $\textsf{best}_f$ for feature combinations $f \in 2^\mathcal{F}$, is chosen as the final 
model, coined \textbf{\textsf{final-M}} in this paper.  By doing so, we  found out that the
\textbf{\textsf{final-M}} is trained using {\it all features except entropy and unalikeability}, reaching  a
accuracy of $80\%$ and $82.003\%$ for \textbf{TestPos} and 
\textbf{TestNeg} datasets, respectively. 

For \textbf{TestPos} and \textbf{TestNeg} datasets separately, Figure~\ref{fig:one-svm} 
compares the performance among a one-class SVM model
built on positive and negative samples separately, \textbf{\textsf{final-M}}, and the average 
performance of all $\textsf{best}_f$ created on feature combinations $f \in 2^\mathcal{F}$.
From the figure, we can observe that the performance of the classification model 
created from interesting training samples using one-class SVM reaches $77.46\%$ 
accuracy for \textbf{TestNeg}. But its performance is very poor (only $7.5\%$) for 
\textbf{TestPos}. This model is clearly unable to detect outliers and is under-fitting the data,
which is unacceptable for a reasonable classifier. Though, the classification model built on 
non-interesting samples using one-class SVM has consistent performance on both 
\textbf{TestPos} and \textbf{TestNeg}, the performance is inferior to the average performance 
of $\textsf{best}_f$ created by using $\nu$-SVM method. Finally, we see that \textbf{\textsf{final-M}} is 
clearly outperforming all other models.

\subsection{Evaluation Based on User Study}  
 
Let us now evaluate  the classification model based on user study
to validate our whole approach. As mentioned earlier in Section~\ref{sec:validation}, we consider different levels of user 
agreement based on majority voting.
For each $x/y$ agreement level, we divide the ground truth into two test files. The samples 
which are marked `interesting' in the ground truth of user study with $x/y$ agreement level,
is denoted as \textbf{x/y-userPos}. On the other hand, the samples which are marked 
`non-interesting' in the ground truth of $x/y$ agreement level are denoted as 
\textbf{x/y-userNeg}. Further, we exclude the samples which are marked as `not sure' from the 
ground truth. As the agreement level decreases, the user-evaluated test dataset contains more 
non-interesting samples than interesting ones. For the lowest agreement label,  the
\textbf{5/9-userNeg} dataset contains almost two times more samples than  dataset
\textbf{5/9-userPos}.

\subsubsection{Performance of Individual Features}
To understand how well each of the features, i.e., the statistical measures, can classify the  
data, Figure~\ref{fig:featurecom} compares the performance across  the $\textsf{best}_f$
classification models created with {\it one single feature} $f \in \mathcal{F}$. The ground 
truth is considered 6/9 agreement level of user assessment for this figure. We can see from the 
figure that the model created based on \textsf{P-Diversity} is
outperforming the other models ($72.41\%$ overall accuracy),
for complete user study. Another model which is also performing well ($70.69\%$ overall 
accuracy) is based on \textsf{Max-Info-gap}. The strong performance of these models is consistent 
throughout all agreement level, except 8/9 agreement level where the model based on entropy 
reaches slightly higher accuracy---not shown here for space limitations.
Figure~\ref{fig:featurecom} also shows that all these models have an inferior precision value
for 6/9-userPos than 6/9-userNeg. 
The precision improves as the agreement level increases and reaches 100\% and 83.33\% 
for 9/9-userPos and 9/9-userNeg respectively.
Comparing the $F_1$-measure, i.e., the harmonic mean of precision and recall, we observe  
that the model created using \textsf{P-Diversity} is outperforming the other models created on single 
features. It supports our claim that the proposed measures can capture the perception of 
of human interests better than the existing ones. Also, the performance of the classification 
model trained on \textsf{Max-Info-gap} shows superiority over the model trained on 
\textsf{Max-Coverage},
as discussed in Section~\ref{sec:novel}. 

\begin{figure}[t]
    \centering    
     \subfigure[\textbf{6/9-userPos}] {\includegraphics[width= 28mm, angle= 270]{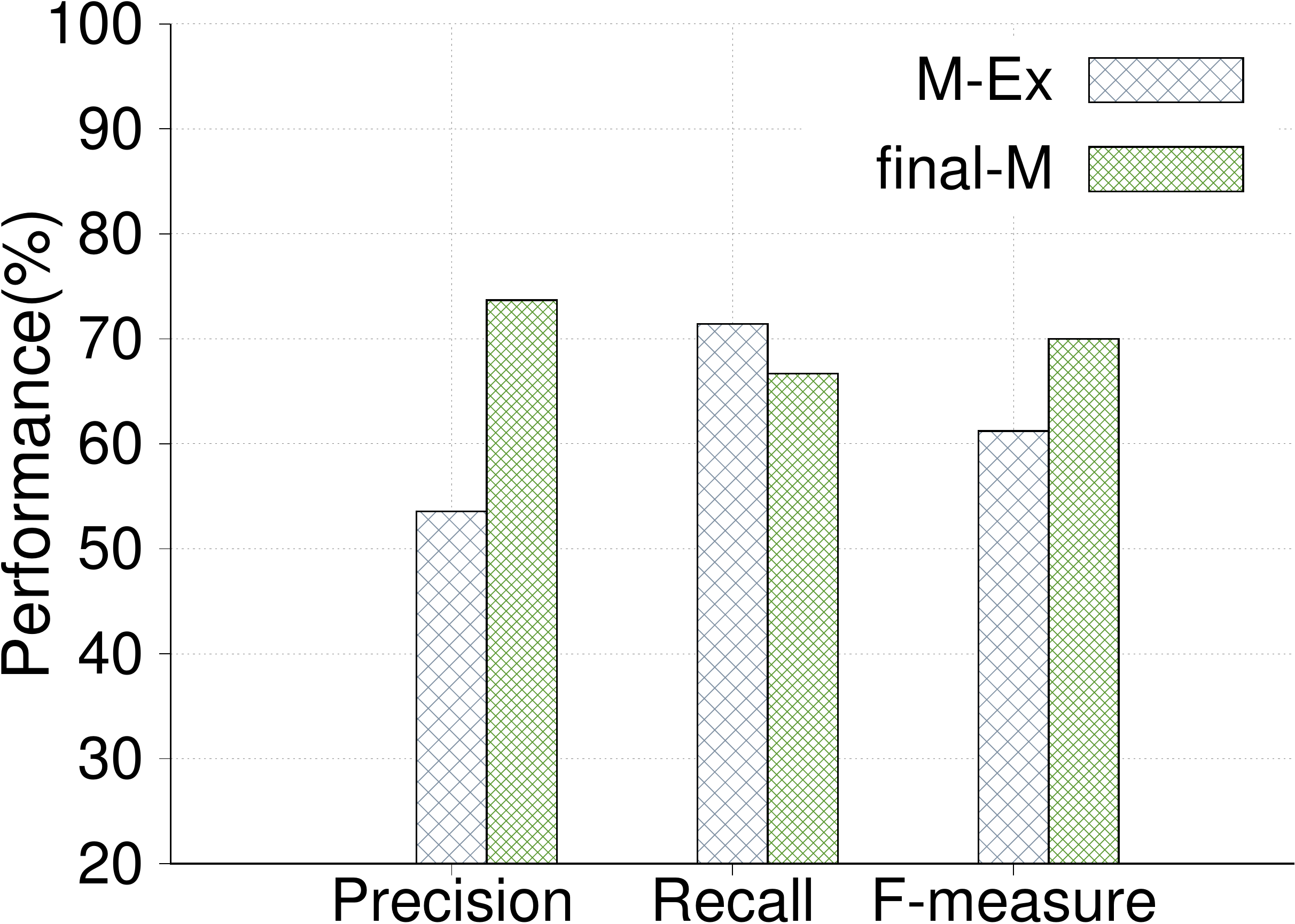}} 
       \subfigure[\textbf{6/9-userNeg}] {\includegraphics[width= 28mm, angle= 270]{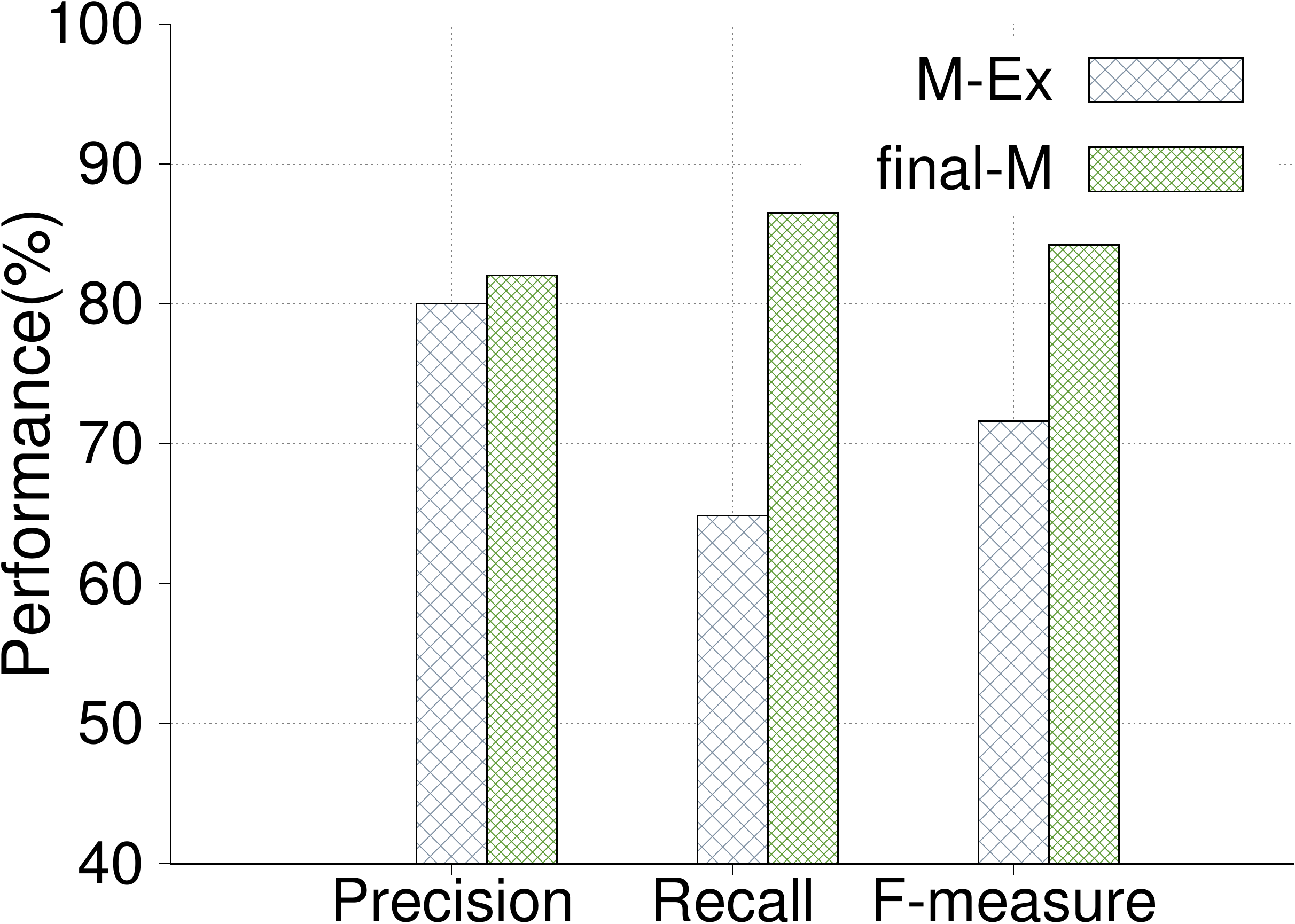}}
       \caption{Performance of \textbf{M} vs. \textbf{M-Ex} }
    \label{fig:model-compare}
\end{figure} 

\subsubsection{Performance of Feature Combinations}
Now we investigate the performance of the \textbf{\textsf{final-M}} which is trained using the
features {\bf \textsf{Peculiarity, P-Diversity, Max-Coverage, Max-Info-Gap, and P-Peculiarity}} 
with the model created based on the existing measures only, i.e., \textsf{Entropy, Unalikability, 
Peculiarity, and Max-Coverage}, denoted as \textbf{\textsf{M-Ex}}. 
\textbf{\textsf{final-M}} achieves $\mathbf{79.31}\%$ overall accuracy, which is much higher 
compared to \textbf{M-Ex} that achieves only $67.24\%$ for the 6/9 agreement level.
Figure~\ref{fig:model-compare} shows that \textbf{\textsf{final-M}} is more robust and achieves 
higher $F_1$-measure than \textbf{\textsf{M-Ex}} based on 6/9 agreement level of user study.

 \begin{table}[t]
\centering
\caption{Performance of \textbf{\textsf{final-M}} for different agreement \\levels}
\label{M-performance}
\small
\begin{tabular}{|@{$\,$}l@{$$}|lll@{$\,$}|lll@{$\,$}|l@{$\,$}|} \hline
\multicolumn{1}{|l@{$\,$}|}{Agreem.} & \multicolumn{3}{c|}{\textbf{userNeg} Samples}  &  \multicolumn{3}{c|}{\textbf{userPos}  samples} & \multicolumn{1}{c|}{} \\  
\ level & Rec. \ \  & Prec. & $F_1$ & Rec. \ \  & Prec. &$F_1$& Acc.   \\  \hline
\ {\bf 9/9} & 100.0 & 83.33 & 90.9 & 83.33 & 100.0 & 90.9 & 90.9 \\ \hline
\ {\bf 8/9} & 91.67 & 78.57 & 84.61 & 70.0 & 87.5 & 77.78 & 81.81  \\ \hline
\ {\bf 7/9} & 86.36 & 82.61 & 84.44 & 75.0 & 80.0 & 77.41 & 81.58 \\ \hline
\ {\bf 6/9} & 86.48 & 82.05 & 84.21 & 66.67 & 73.68 & 70.0 & 79.31 \\ \hline
\ {\bf 5/9} & 84.21 & 78.69 & 81.36 & 53.57 & 62.5 & 57.69 & 74.12  \\ \hline
\end{tabular}
\end{table}

Table~\ref{M-performance} reports on the performance of \textbf{\textsf{final-M}} for different 
agreement levels. From the table we can see that the classification performance of the model 
increases as agreement level increases. This is expected, as larger agreement means that 
human evaluators had no difficulty to accomplish the classification task in a consistent way,
indicating that the task is relatively easy to solve. For such presumably simple tasks, 
the successful classification by automated means is, thus, also more likely. 

\begin{figure}[t]
    \centering    
   \includegraphics[height=0.85\columnwidth, angle= 270]{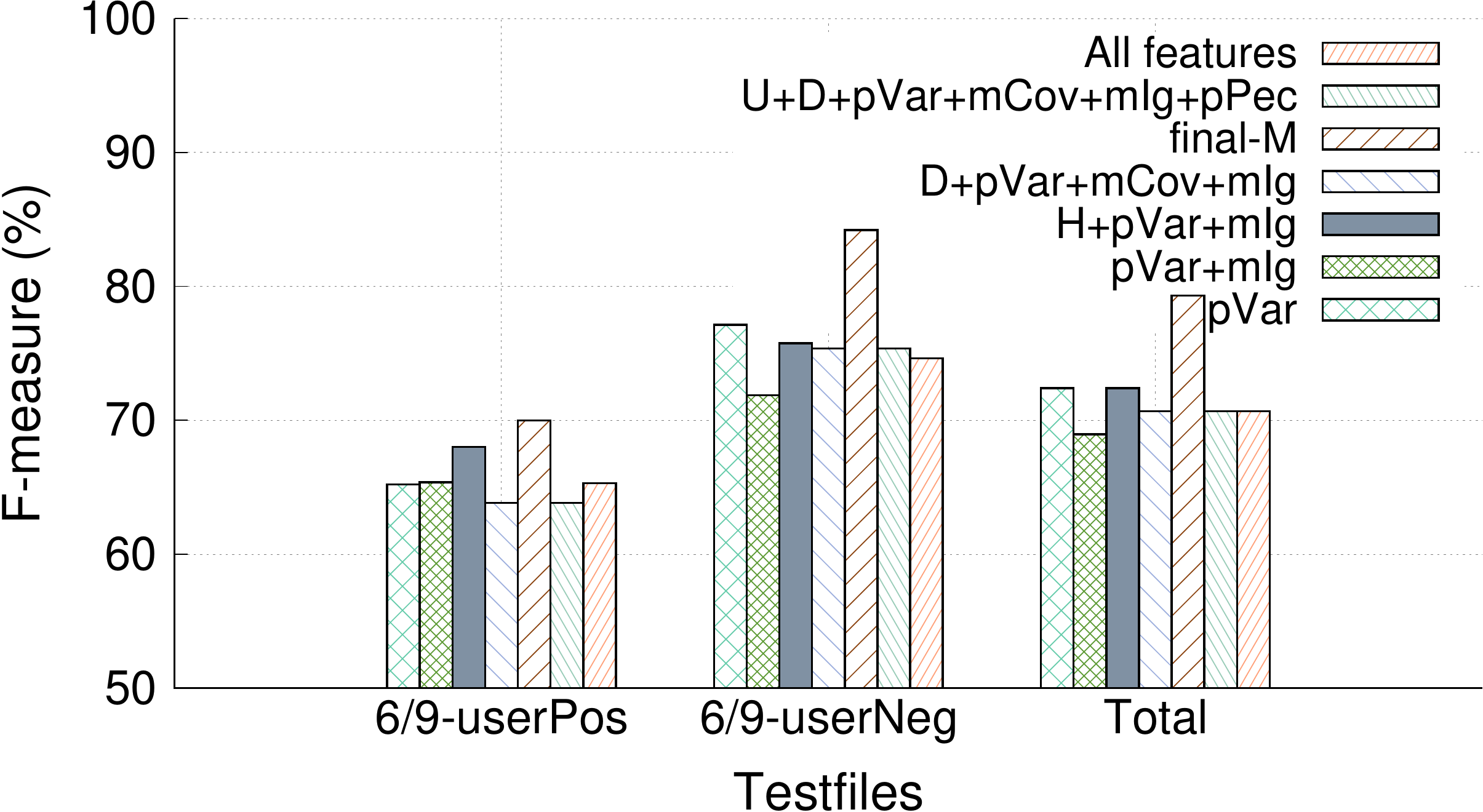}
     \caption{Comparison among best performing models for different numbers of used features, for 6/9 agreement level.}
    \label{fig:fcombi-study}
\end{figure}  

Next, we categorize the  $\textsf{best}_f$ model we have created earlier 
from all possible feature combinations into 7 groups based on the number of features are used 
to train the model, i.e., $1\leq |f| \leq  |\mathcal{F}|$. Then, the best performing model from 
the each group is taken and their performances is reported in Figure~\ref{fig:fcombi-study} 
based on 6/9 agreement level. In this figure we also mention which features are used to create the 
best one among the group. For instance, the model using the features Entropy, P-Diversity, and Max-Info-Gap is performing best among all the models created by combining any 3 features from $\mathcal{F}$. 

\subsubsection{Assessment of Main Hypothesis}
\begin{figure}[t]
    \centering    
     \subfigure[Interesting samples] {\includegraphics[width= 28mm, angle= 270]{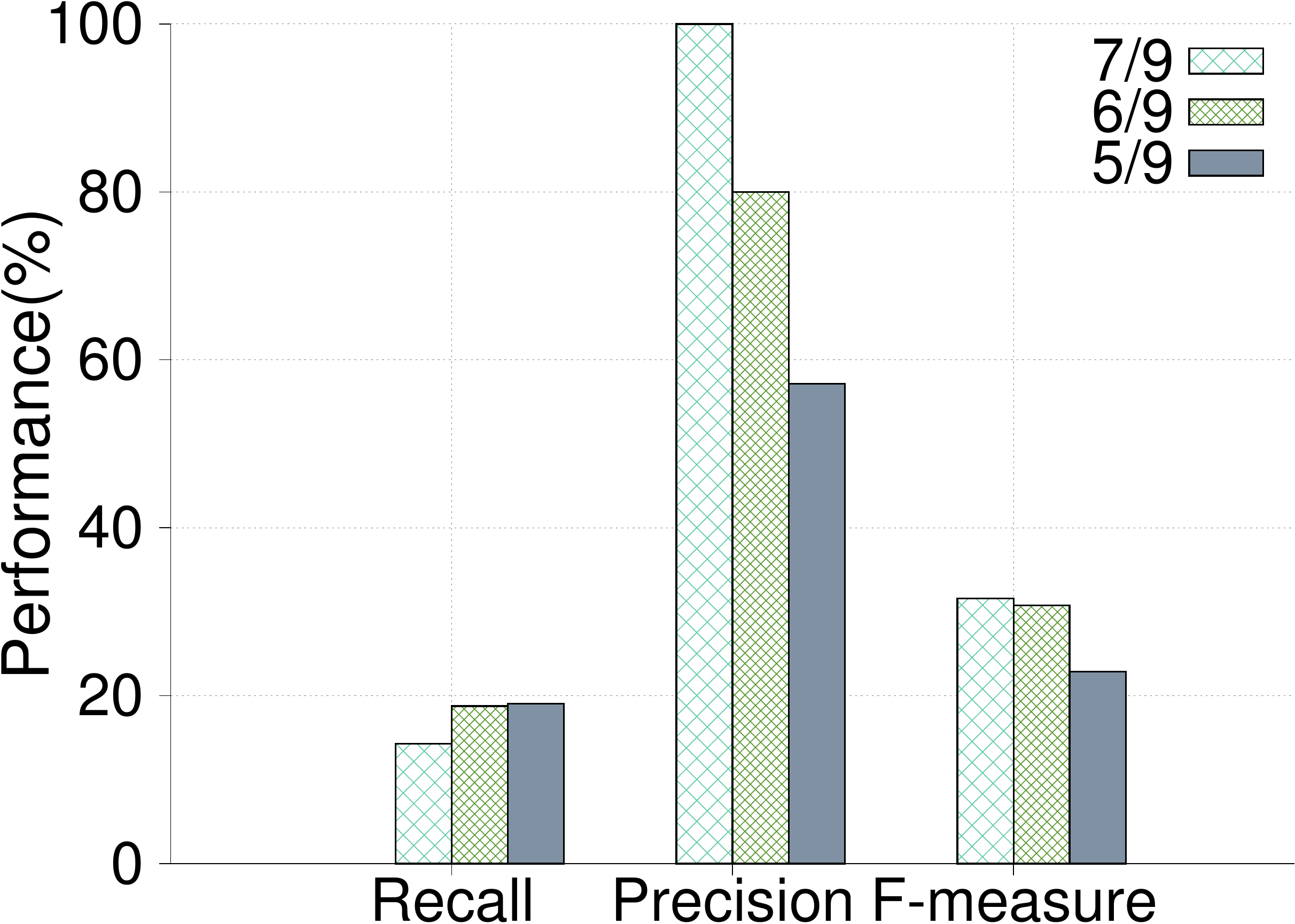}}
       \subfigure[Non-interesting samples] {\includegraphics[width= 28mm, angle= 270]{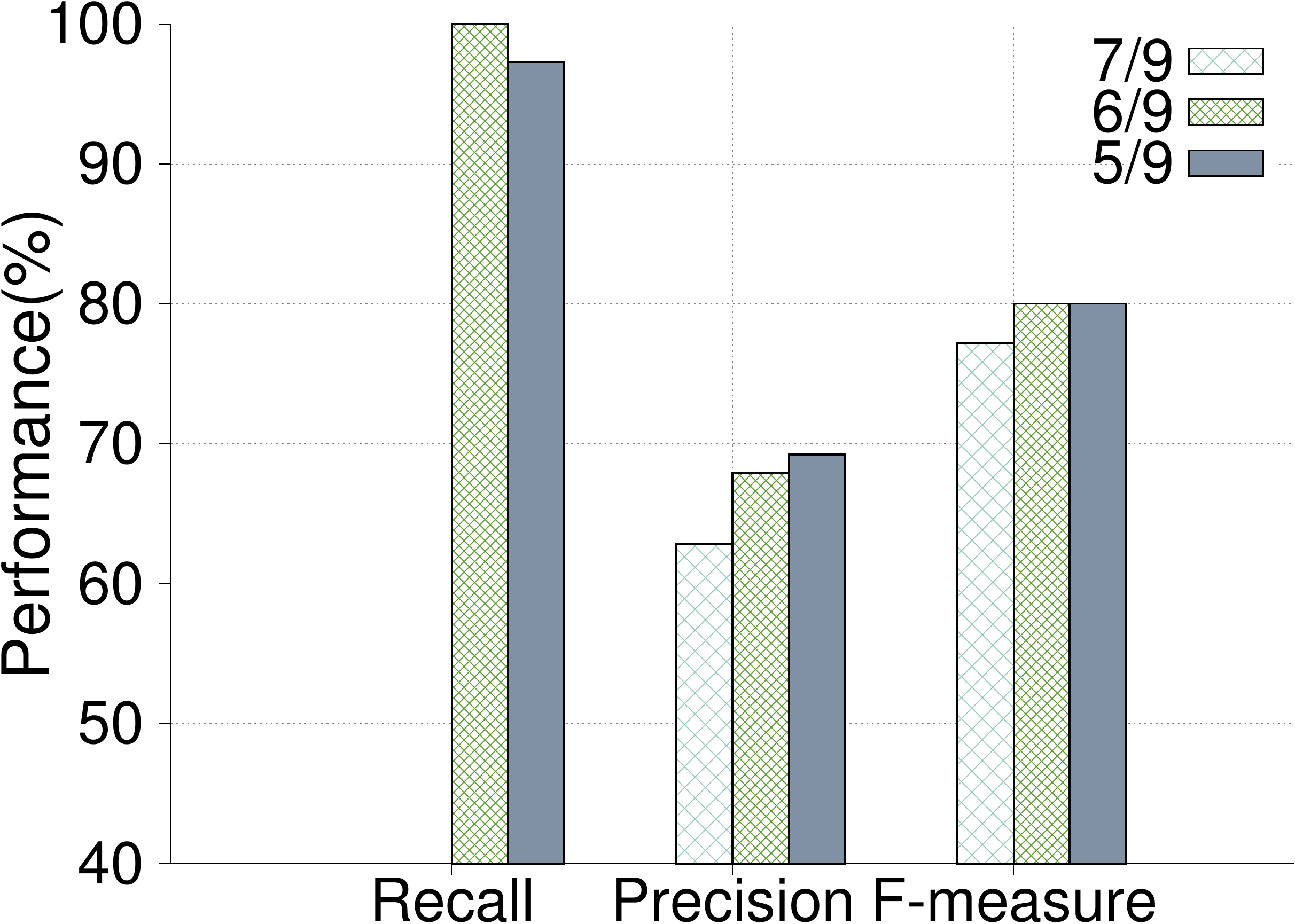}} 
       \caption{Performance of Algorithm~\ref{alg:1}}
    \label{fig:wiki-compare}
\end{figure}

Figure~\ref{fig:wiki-compare} presents an evaluation of Algorithm~\ref{alg:1},  respectively our main hypothesis, directly comparing 
the labels retrieved by the algorithm for \textbf{userPos} and \textbf{userNeg} dataset with the 
human-labeled ground truth for different agreement levels. 
Figure~\ref{fig:wiki-compare} shows that the algorithm can identify the positive sample 
precisely but missed out 
many samples that are marked as `interesting' by users which is reflected by low recall and high
precision on \textbf{userPos} data, shown in Figure~\ref{fig:wiki-compare}.
We believe, one reason behind this low performance on \textbf{userPos} data is  the incompleteness 
of Wiki\-pedia,  discussed earlier in Section~\ref{sec:createdata}. 
Figure~\ref{fig:wiki-compare} also shows that for \textbf{userNeg}, our proposed 
algorithm almost correctly identifies all samples that are marked as `non-interesting' by users. 
Moreover, \textbf{\textsf{final-M}} which is created based on the training samples retrieved by 
Algorithm~\ref{alg:1} reaches reasonable performance for both \textbf{userPos} and 
\textbf{userNeg} data as presented in Table~\ref{M-performance}. 
{\it These findings strongly  support  our working hypothesis} (cf., Section~\ref{sec:createdata}) which states 
that positive and negative training samples can be derived from the presence, respectively 
absence, of tables in Wiki\-pedia. 

\begin{figure}[t]
    \centering    
   \includegraphics[height=0.85\columnwidth, angle= 270]{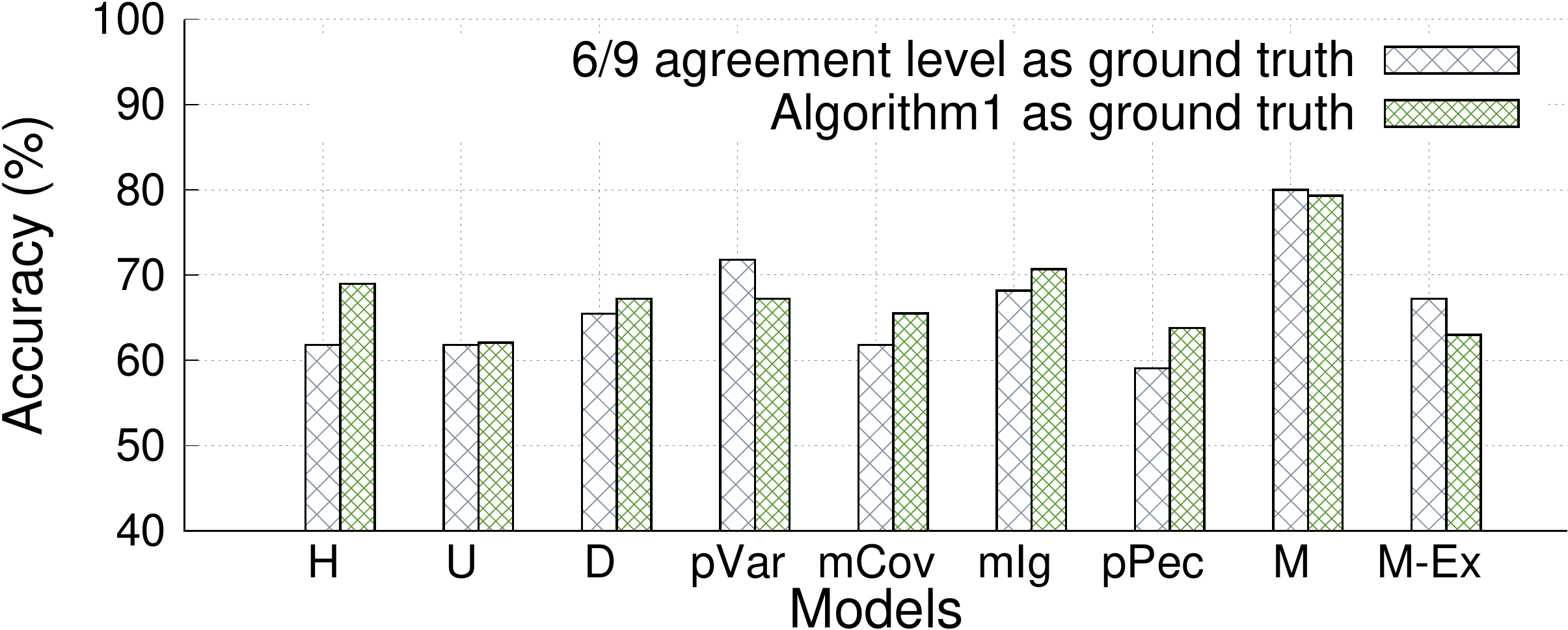}
     \caption{Performance of Models w.r.t. different Ground Truths.}
    \label{fig:study1}
\end{figure}

Figure~\ref{fig:study1} shows that the models created on different features achieve almost the same classification accuracy
for user-study samples according to two different ground truth definition: {\it (i)} the level retrieved by Algorithm~\ref{alg:1}
and {\it (ii)} 6/9 agreement level from our user study. 
This performance remains consistent for all other models which is not shown in the figure. It also emphasizes the robustness of our classification model.

\subsection{Lesson Learned}

We can summarize the main findings of the experimental studies in the following {\bf four lessons} learned.

\begin{enumerate}
\item Our working hypothesis, which is the main idea behind Algorithm~\ref{alg:1},
is well-grounded: 
The positive and negative training samples in the obtained training data are generally confirmed by human evaluators.
\item Our study shows that \textsf{P-Diversity} outperforms existing measures for the task to capture diversity in our context. \textsf{Max-Info-Gap} performs better than \textsf{Max-Coverage} and \textsf{Entropy}.
\item \textbf{\textsf{Final-M}} is able to achieve 79.31\% overall accuracy, with ground truth 
given by the 6/9 user-agreement level. It uses all the features discussed in this paper, except entropy and unalikeability, to learn the model.
\item \textbf{\textsf{Final-M}}  can accurately classify the data even when disagreement among the users increases (i.e., the classification task gets more difficult).
\end{enumerate}


\section{Related Work}
\label{sec:relatedworks}

Understanding  data is a primary necessity for scientific discovery. Data analysts  often use OLAP-cubes~\cite{DBLP:journals/datamine/GrayCBLRVPP97, DBLP:conf/edbt/SarawagiAM98} or mining 
algorithms~\cite{DBLP:journals/kais/WuKQGYMMNLYZSHS08} to explore data. Different approaches to drill-down
 operations are proposed in literature to make OLAP smarter 
and more efficient for large data. Joglekar et al.~\cite{DBLP:conf/icde/JoglekarGP16} propose an interaction 
operator to extend the scope of drill-down operations. It allows online user interaction  and enables
browsing the top-k most interesting explorative facts about the data, based on the dimensions preferred by the
data analyst. Our approach can be used orthogonally, as an enabling step, to such approaches by 
providing recommendations of meaningful dimensions.
Algorithms for clustering and classification~\cite{DBLP:journals/kais/WuKQGYMMNLYZSHS08} commonly 
define the interestingness of discovered patterns in statistical ways, without considering the users' utility. 
Bie~\cite{DBLP:conf/ida/Bie13} proposed a mathematical framework to formulate 
interestingness of mining patterns,  considering user utility as one important parameter. 
Tuzhilin~\cite{Tuzhilin95onsubjective} proposes an approach to measure interestingness w.r.t.   
the belief system of users. All these works need user intervention to capture the utility function. 

The survey by Geng and Hamilton~\cite{Geng:2006:IMD:1132960.1132963} 
provides a detailed discussion on subjective and objective measures used to capture 
`interestingness' of data for association or classification rule mining.  
Henzgen and H\"ullermeier~\cite{DBLP:conf/dis/HenzgenH14} present an analogy of the itemset-mining 
measures support and interest applied to mining subrankings.  
Different context-specific diversity measures are proposed in the area of Web 
search~\cite{DBLP:conf/www/RafieiBS10, DBLP:conf/wsdm/AgrawalGHI09, DBLP:conf/sigir/DangC12}, 
entity summarization~\cite{DBLP:journals/jiis/SydowPS13},  and recommender systems~\cite{DBLP:conf/sigir/SchedlH15}. 
These diversity measures are not applicable to our work and not all of them are based on empirical 
probabilities. Hilderman and Hamilton~\cite{Hilderman01evaluationof} present a
comparative evaluation of diversity measures that are available to capture interestingness of patterns. 
They show that a small subset of these measures are in fact useful and can capture the 
characteristics of interestingness. In this work, we have used the diversity measures, Peculiarity 
and Unalikeability, based on empirical probabilities, discussed by Kader and Perry~\cite{journals/Gary07}.

Web tables are considered a rich source of information and mining Web tables is very popular nowadays for
extracting accurate and robust information~\cite{DBLP:conf/wsdm/CrestanP11, DBLP:conf/er/WangWWZ12, DBLP:conf/kdd/SarawagiC14}. 
In particular, tables in Wikipedia are considered a rich and credible source for information, leading to knowledge bases like YAGO and DBpedia that are created based on Wiki\-pedia pages.
Availability of such high quality information in Wiki\-pedia facilitate mining of Wiki\-pedia tables to explore knowledge about entities ~\cite{DBLP:conf/cikm/IbrahimRW16, DBLP:conf/kdd/BhagavatulaND13, DBLP:journals/pvldb/ChirigatiLKWYZ16}. 

The concept of SVMs is a well-known supervised learning algorithm and generally considered  robust and accurate.
Different variations of SVM are developed in order to cope with different characteristics
of the available training data. Sch\"olkopf et al.~\cite{DBLP:journals/neco/ScholkopfPSSW01} 
 present the {\it one-class SVM} which is able to learn  a
classification model  that is presenting only a single class. This SVM variant is shown to
perform well in cases where the training data is unbalanced.
 Another variant, coined $\nu$-SVM, 
is discussed in~\cite{DBLP:journals/neco/ScholkopfSWB00,Chen05atutorial} for learning classification model from noisy data. 
SVM is also capable to classify non-linear data by using various kernel methods, discussed 
in~\cite{Scholkopf:2001:LKS:559923}. One of the widely used  kernel methods is the Gaussian 
kernel~\cite{DBLP:journals/jmlr/ChangHCRL10}.


\section{Conclusion}
\label{sec:conclusion}
Categorical attributes allow grouping items in meaningful ways and, thus, are key to render data accessible. 
We presented a new approach to capture human interest in categorical attributes.
We started with providing the  hypothesis that training data can be derived from Wikipedia based on the presence or absence of specific tables.
We motivated and defined three new statistical measures to capture subjective interestingness measures for our context. 
The results of the experimental study, involving a user study, show that using features combination, a classification model can  
well reflect human interests on categorical attributes. It also shows that the proposed statistical measures 
are more suitable to capture the characteristics of the interesting categorical attributes compared to the traditional measures like information entropy.
Applications are manifold, ranging from mining interesting patterns in data or understanding dependencies between table columns, to clustering or filtering entities and  finding dimensions of interest in  data warehouses.



\end{document}